\documentclass{mindlab}

\usepackage{microtype}
\usepackage{hyperref}
\usepackage{amsmath}
\usepackage{subcaption}  % 放在导言区
\usepackage{url}
\usepackage{booktabs}
\usepackage{graphicx} 
\usepackage{wrapfig}
\usepackage{simpleicons}
\usepackage{xcolor}
\usepackage{listings}
\lstdefinestyle{a2uiprompt}{
  basicstyle=\ttfamily\small,
  breaklines=true,
  columns=fullflexible,
  backgroundcolor=\color{gray!10},
  frame=single,
  framerule=0pt
}
\lstnewenvironment{minted}[2][]{\lstset{style=a2uiprompt}}{}
\DeclareRobustCommand{\huggingfaceicon}{{\normalfont\simpleicon{huggingface}}}
\DeclareRobustCommand{\githubicon}{{\normalfont\simpleicon{github}}}

\title{Macaron-A2UI: A Model for Generative UI in Personal Agents}
\author{Fancy Kong}
\author{Congjie Zheng}
\author{Murphy Zhuang}
\author{Rio Yang}
\author{Sueky Zhang}
\author{Hao Fu}
\author{Gene Jin}
\author{Song Cao}
\author{Kaijie Chen}
\author{Andrew Chen}
\author{Pony Ma}
\affiliation{\small{Mind Lab}}
\metadata[\huggingfaceicon\ Models]{\url{https://huggingface.co/collections/mindlab-research/macaron-a2ui}}
\metadata[\githubicon\ Benchmark]{\url{https://github.com/MindLab-Research/Macaron-A2UI-Bench}}
\metadata[\faEnvelope\ Correspondence]{\texttt{\textcolor{mindlabblue}{\{fancy, andrew, pony\}@mindlab.ltd}}}
\metadata[\faCalendar\ Date]{May 2026}

\abstract{As personal agents evolve to handle complex, user-centric tasks, static plain-text chat is rapidly becoming a bottleneck. Generative UI emerges as the necessary new interface layer, dynamically synthesizing the right controls, options, and state from the interaction context in real time. We present Macaron-A2UI, a model for Generative UI in personal agents. Our goal is to move beyond text-only interaction by enabling agents to generate natural language together with lightweight, executable UI actions for information collection, preference refinement, confirmation, and multi-goal organization. We build a large-scale Generative UI corpus from heterogeneous dialogue sources, introduce A2UI-Bench for controlled evaluation, and train 30B, 235B and 754B models with parameter-efficient LoRA-based supervised fine-tuning followed by reward-driven reinforcement learning. 
The best Macaron-A2UI model reaches 75.6 overall on A2UI-Bench without explicit schema hints, surpassing the strongest full-schema frontier baseline. We release the models, benchmark, and evaluation protocol to support future work on Generative UI for personal agents.}

\begin{document}

\maketitle

\section{Introduction}

Powerful AI agents and code-generating models are changing the core assumptions behind software interfaces. Human-computer interaction no longer relies solely on fixed screens designed for broad populations~\citep{findlater2004comparison}. Interfaces are instead becoming flexible and personalized. They can be created at the moment of interaction to match the user's goal, context, and next actions~\citep{gajos2010automatically}. This shift establishes Generative UI as an essential direction for future software development. The agent can create an appropriate interaction surface when plain text is insufficient~\citep{todi2021adapting}.

This flexibility materializes as executable interfaces generated within the interaction loop. These interfaces prove vital for tasks requiring structured interaction~\citep{cohen2019foundations}. Users often must provide information, compare options, confirm decisions, or organize multiple goals in a single turn~\citep{norman1986cognitive, qian2025userbench, budzianowski2018multiwoz, kong2026infopo}. Here, long text replies slow reading and increase cognitive load. Lightweight generative interfaces address this directly, making complex interactions shorter, clearer, and easier to complete~\citep{avula2022effects}.

\begin{figure}[t!]
    \centering
    \includegraphics[width=0.95\linewidth]{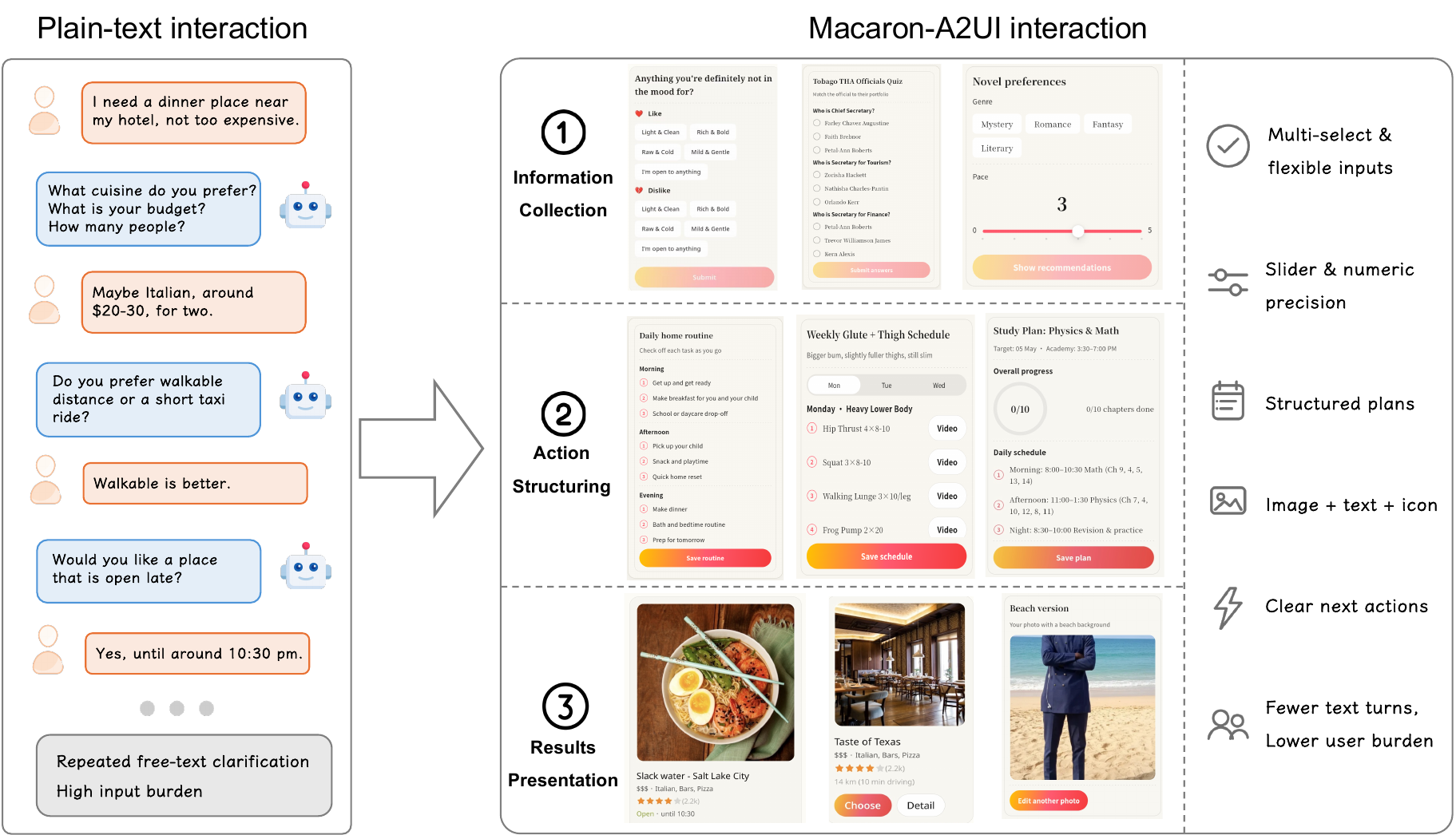}
    \caption{Many dialogue turns that are cumbersome in plain text become more efficient when the assistant can render lightweight structured interfaces.}
    \label{fig:intro}
\end{figure}

While modern language models readily produce structured outputs~\citep{dong2025protod, geng2025generating, patil2025berkeley}, Generative UI for personal agents remains underexplored as a complete learning problem~\citep{chen2025generative, liu2026alignui}. Current research primarily focuses on plain-text dialogue, code generation~\citep{wu2024uicoder, si2025design2code, wu2026autowebworld}, or navigating existing interfaces~\citep{zhao2025worldgui, wu2025gui}. A unified formulation of agent-side UI generation is still missing. Specifically, the field lacks large-scale UI-grounded dialogue supervision, evaluation benchmarks that separate protocol validity from interaction quality, and evidence that models can internalize this capability without relying on long schema prompts.

% Recent progress in language models has made this direction more realistic. Dialogue research has increasingly moved beyond purely reactive, text-only response generation toward mixed-initiative, proactive, and collaborative interaction, where the assistant may need to ask clarifying questions, steer the interaction, or externalize intermediate structure rather than only reply in free text (Cohen, 2019; Wu et al., 2023; Nie et al., 2024; Dong et al., 2025; Yoon et al., 2025). At the same time, recent studies in conversational search and dialogue evaluation suggest that complex multi-turn interactions are poorly captured by simple one-shot answer quality, and instead require more fine-grained assessment of interaction structure, initiative, and turn-level effectiveness (Kelly et al., 2022; Mo et al., 2025; Acikgoz et al., 2025). Still, multi-turn conversational UI generation has not yet been established as a complete learning problem. Existing work usually focuses either on plain-text dialogue, interface or webpage synthesis from instructions or screenshots, or agents that act over already existing interfaces (Si et al., 2025; Zhao et al., 2025). These settings leave a gap for assistant-side interface generation under a fixed declarative protocol, especially when the goal is to unify natural dialogue, executable structure, and stable evaluation in a single framework.

In this paper, we study Generative UI for personal agents as a learning problem. Given a system instruction, dialogue history, and the current user message, the model produces a unified response containing natural language and an executable UI action sequence. We instantiate this interface using A2UI, a declarative UI protocol~\citep{a2ui_v08}. This provides a renderable, automatically checkable foundation to rigorously study when an agent should generate a UI, its structural content, and its overall utility.

To support this formulation, we convert four heterogeneous dialogue sources into a Generative UI corpus of over 14,000 samples using a hybrid rule-and-LLM approach with deterministic validation. And we introduce A2UI-Bench which evaluates protocol validity, interaction quality and visual metrics.

Using this setup, we train Generative UI assistants through a parameter-efficient two-stage recipe. LoRA-based supervised fine-tuning establishes text-UI grounding, and reinforcement learning subsequently improves executable interaction quality. This approach targets the minimal-prompt regime, requiring models to internalize UI generation directly. Experiments across model scales demonstrate our pipeline's effectiveness.
Notably, our best model achieves an overall score of 75.6, surpassing the strongest full-prompt frontier baseline, confirming that Generative UI competence can be successfully internalized.

Our work makes three contributions. 

\begin{itemize}
    \item First, we introduce a scalable pipeline for transforming heterogeneous dialogue corpora into multi-turn Generative UI interaction data, combining LLM-based UI annotation with rule-based repair and validation for renderability.
    \item Second, we establish a benchmark for Generative UI interaction modeling, with three task families and a three-level evaluation framework that measures protocol validity, task progression, and user experience.
    \item Third, we develop a parameter-efficient two-stage training recipe, combining LoRA-based schema-light SFT and reward-driven RL, showing that executable UI generation can be internalized without long schema prompts at inference time.
\end{itemize}

\section{Related Works}

Recent work has increasingly explored interface generation as a native capability of foundation models. Generative UI~\citep{leviathan2025generative} shows that LLMs can synthesize rich task-specific interfaces rather than only return linear text, while~\citet{chen2025generative} further argues that proactively generated interfaces can improve interaction quality for information-dense tasks and evaluates them along functional, interactive, and emotional dimensions. AlignUI~\citep{liu2026alignui} incorporates user preferences into the interface design process. More broadly, lots of work studies interface generation from text instructions, design specifications, screenshots, or interaction requirements, including dynamic GUI generation in chat settings, text-to-UI code generation, screenshot-to-code generation, accessibility-aware interface generation, and UX-oriented generative design systems~\citep{wu2024uicoder, si2025design2code, yoon2025a11yn, chen2025genui}. Compared with this line, our focus is not unconstrained webpage or code synthesis, but structured turn-level Generative UI under a fixed declarative protocol and executable rendering constraints.

A second line of research studies agents that operate over existing digital interfaces, spanning web browsing, screen-grounded GUI interaction, and scalable task or trajectory construction. On the web side, recent works evaluate persistent and increasingly multimodal browsing ability~\citep{zhang2026browsecomp, li2025mm}. For screen- and GUI-grounded agents, WorldGUI~\citep{zhao2025worldgui} studies desktop GUI automation under diverse initial states, GUI-Actor~\cite{wu2025gui} advances coordinate-free visual grounding for GUI actions, and recent mobile benchmarks further emphasize ambiguous, proactive, and personalized interaction settings~\citep{sun2026ambibench, yang2025fingertip}. In parallel, recent efforts have begun to explore scalable construction of agent data: some target broader agentic task synthesis, such as TaskCraft~\citep{shi2025taskcraft}, while others are more tightly coupled with GUI settings, including OS-Genesis for reverse task synthesis, GUI-360~\citep{mu2025gui} for large-scale trajectory collection and benchmarking, and Log2Plan~\citep{lee2025log2plan} for adaptive GUI automation with task mining from user behavior logs. In contrast to these lines of work, we focus on assistant-side Generative UI rather than action execution over an existing interface.

% This includes web-agent benchmarks built around browser interaction and realistic online tasks, such as Mind2Web, WebArena, VisualWebArena, and WebLINX, as well as screen- and GUI-grounded agents for mobile and desktop interfaces, such as Android in the Wild, ScreenAI, and ShowUI. More recently, researchers have also explored scalable task and trajectory construction for such agents: while TaskCraft addresses general agentic task synthesis, OS-Genesis, GUI-360, and Log2Plan move more directly toward GUI-grounded trajectory generation, multimodal supervision, and adaptive automation. Different from this line of work, we do not study agent execution over external interfaces, but assistant-side generation of structured conversational UI.

\section{Problem Formulation and A2UI Primer}
\label{sec:problem}

We study A2UI-based Generative UI: given a system instruction, a dialogue history, and the current user message, the model must produce a unified assistant response that contains natural language and, when appropriate, a structured A2UI message sequence. Unlike approaches that ask the model to generate HTML, JavaScript, or framework-specific code, A2UI is a declarative protocol in which the model emits structured messages, and the client renders them using a trusted component catalog. This separation is important for our setting: it makes UI generation safer, more portable across rendering environments, and easier to validate automatically. In this paper, we instantiate all data construction, rendering, and evaluation against A2UI v0.8, the current stable public version.

At a high level, A2UI v0.8 organizes interaction through four message types. \texttt{surfaceUpdate} defines or updates UI components, \texttt{dataModelUpdate} updates application state, \texttt{beginRendering} signals the client to render a surface, and \texttt{deleteSurface} removes an existing surface. 

There are several challenges in A2UI generation. The first one is \textbf{protocol validity}: an output may be syntactically well-formed JSON yet still violate message, reference, typing, or renderability constraints. The second is \textbf{interaction construction}: even a protocol-valid UI may still use the wrong widget type, fail to ground visible choices in the assistant text, or mishandle state updates across turns. The third challenge is \textbf{user-facing quality}: an output may be structurally correct and functionally plausible, yet still add little value beyond plain text, feel abrupt in context, or impose unnecessary cognitive load on the user.

This decomposition directly motivates the rest of the paper. In Section~\ref{sec:data}, we construct an A2UI-grounded corpus that teaches the model how to generate protocol-compliant UI. In Section~\ref{sec:benchmark}, we design a benchmark that evaluates model's ability when facing these challenges. In Section~\ref{sec:exp}, we show that learning under this formulation benefits from a two-stage training pipeline: supervised fine-tuning first stabilizes the response format and basic text--UI grounding, while reinforcement learning further improves the quality of executable interaction.
% \section{Dataset}

% \paragraph{Source corpora and coverage.}
% SGD, MultiWoz, Annomi, Esconv

% \paragraph{LLM-assisted annotation pipeline}
% Determine whether the current round is suitable for UI development;
% If suitable, determine the UI type;

% Complete fields and semantic constraints;

% Map the original natural language interaction into A2UI expressions.

% \paragraph{Rule-based validation and repair}
% schema validation, component whitelist validation, field normalization, surface/action repair, illegal output rollback, etc

% \paragraph{Dataset statistics and quality checks}

% Suggested packages:
% \usepackage{booktabs}
% \usepackage{graphicx}
% \usepackage{multirow}
% \usepackage{array}

\section{A2UI Corpus Construction}
\label{sec:data}

\begin{figure}[t]
    \centering
    \includegraphics[width=\textwidth]{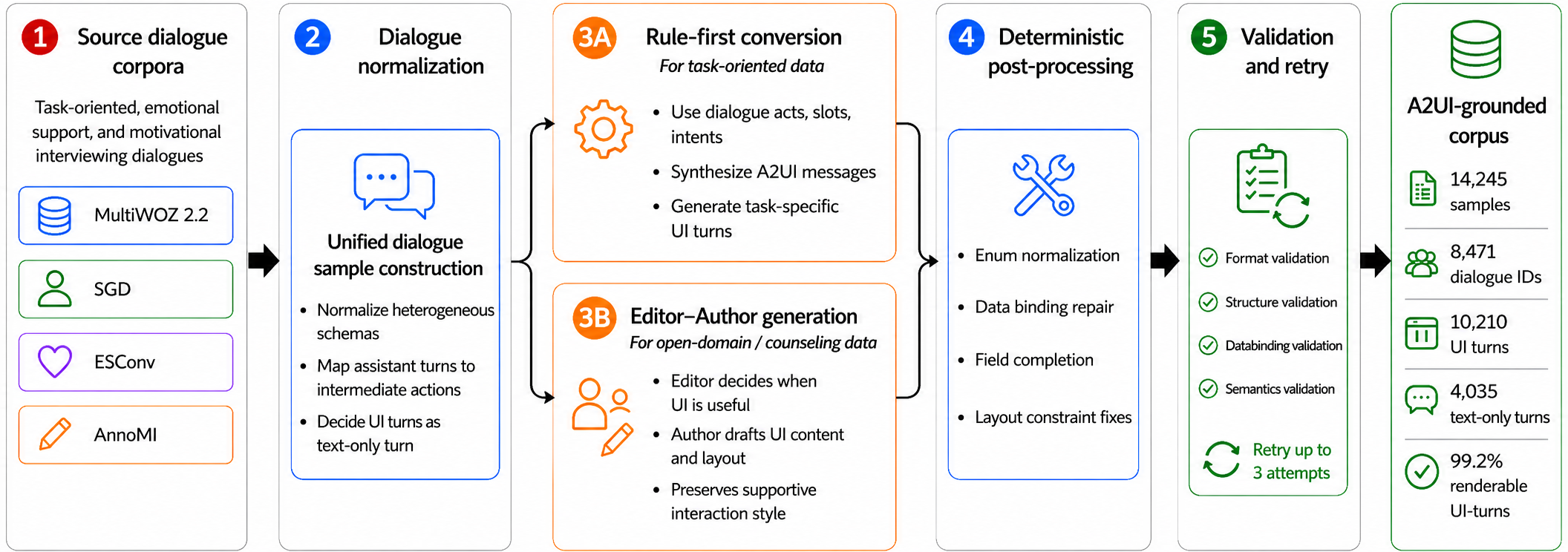}
    \caption{Overview of the A2UI corpus construction pipeline. Source dialogues are normalized and mapped to intermediate interaction actions; task-oriented data are converted with rule-first generators, open-domain data are annotated through the Editor--Author process, and all outputs pass deterministic post-processing, validation, and retry before entering the final corpus.}
    \label{fig:data-pipeline}
\end{figure}

We construct an A2UI-grounded dialogue corpus from four heterogeneous source datasets: MultiWOZ~2.2, Schema-Guided Dialogue (SGD), ESConv, and AnnoMI. Our goal is not only to attach valid A2UI payloads to existing conversations, but to build training data that teaches a model three behaviors simultaneously: when to produce UI, what UI to produce, and how to produce protocol-compliant UI under lightweight prompting.

\subsection{Source Corpora and A2UI-Oriented Sample Construction}
\label{subsec:data-source}

We begin with four dialogue corpora that cover complementary interaction regimes: task-oriented assistance (MultiWOZ and SGD), emotional support (ESConv), and motivational interviewing (AnnoMI). These sources differ substantially in annotation schema, dialogue length, and interaction styles, thus requiring normalization to a unified sample format.

\paragraph{Basic unit of supervision.}
A \emph{dialogue} denotes a sampled source conversation segment. A \emph{training sample} is a $(\texttt{context}, \texttt{response})$ pair, where \texttt{context} contains the full dialogue history up to the current assistant turn, and \texttt{response} contains the assistant's natural-language reply together with an optional A2UI payload. One dialogue can therefore yield multiple training samples, one per assistant response. Following the rest of the paper, we use \emph{turn} to refer to such a training sample. A \emph{UI-turn} is a sample whose response contains a non-empty A2UI message. A \emph{text-only turn} contains only natural language and no A2UI.

\paragraph{Dialogue normalization.}
Before A2UI annotation, we merge consecutive utterances from the same speaker to obtain strict user--assistant alternation. This step removes dataset-specific segmentation artifacts and yields a consistent dialogue history format across all four sources.

\paragraph{Unified intermediate representation.}
To bridge heterogeneous source annotations, we map dataset-specific signals into a compact intermediate interaction representation. For MultiWOZ and SGD, dialogue acts, intents, and slot annotations are mapped to actions such as collecting missing constraints, presenting options, and confirming a decision. For ESConv and AnnoMI, support strategies and counseling behaviors are mapped to interaction patterns such as guided selection, reflection support, confidence elicitation, or action planning. We then map these intermediate actions to A2UI component families. For example, categorical choices are mapped to selection widgets, numeric or ordinal values to slider-style controls, boolean fields to check boxes, and temporal arguments to date/time inputs.

Table~\ref{tab:corpus_composition} summarizes the resulting corpus composition. In total,
we sample 4,306 base dialogues and obtain 14,245 assistant-turn training samples, including
10,080 original samples and 4,165 component-targeted augmented samples. Two dataset-specific
decisions are worth noting. First, for AnnoMI, we retain only the high-quality subset and then
expand it through component-targeted augmentation, which increases coverage of counseling
and motivational-interviewing interaction patterns. Second, for SGD, we use a single-turn
sampling strategy for most dialogues, selecting one highly informative assistant turn per
dialogue in order to maximize service coverage rather than dialogue depth.

\begin{table}[t]
\centering
\small
\setlength{\tabcolsep}{4pt}
\begin{tabular}{llrrrrr}
\toprule
Source & Domain & Base dlg. & Orig. / Aug. & Samples & UI / Text & UI ratio \\
\midrule
MultiWOZ & Task-oriented & 997 & 3,673 / 1,751 & 5,424 & 4,361 / 1,063 & 80.4\% \\
SGD & Task-oriented & 3,109 & 3,692 / 1,065 & 4,757 & 3,761 / 996 & 79.1\% \\
ESConv & Emotional support & 100 & 760 / 338 & 1,098 & 604 / 494 & 55.0\% \\
AnnoMI & Motiv. interview. & 100 & 1,955 / 1,011 & 2,966 & 1,484 / 1,482 & 50.0\% \\
\midrule
Total & -- & 4,306 & 10,080 / 4,165 & 14,245 & 10,210 / 4,035 & 71.7\% \\
\bottomrule
\end{tabular}
\caption{
Corpus composition by source. Base dlg. counts sampled source dialogues before augmentation.
Orig. / Aug. denotes assistant-turn training samples from base dialogues and component-targeted augmentation.
UI / Text denotes samples with and without A2UI payloads.
}
\label{tab:corpus_composition}
\end{table}

\subsection{Hybrid A2UI Annotation and Augmentation Pipeline}
\label{subsec:data-pipeline}

We construct A2UI responses with a hybrid rule-and-LLM pipeline. The core design principle is to use deterministic structure whenever source annotations already constrain the interaction semantics, and to mainly use LLMs where the source dialogue leaves UI decisions under-specified.

\paragraph{Task-oriented data.}
For MultiWOZ and SGD, UI generation is primarily rule-driven. We use a state-machine-style conversion process that tracks surface lifecycle events such as creation, update, and removal across turns. The source annotations determine what information is missing, what options are available, and what confirmation or correction step is needed next. The generator then instantiates the corresponding A2UI surfaces and widgets. In this regime, LLMs are used mainly to rewrite or polish user-facing text so that the final responses read naturally while preserving the original dialogue semantics.

\paragraph{Open-domain data.}
For ESConv and AnnoMI, source annotations do not directly specify a concrete UI. We therefore use a two-stage LLM process. An \emph{Editor} pass first plans the dialogue globally, deciding which turns should contain UI and what interaction type is appropriate. An \emph{Author} pass then generates the local component content for each selected turn, including widget text, option semantics, and layout-level organization. This decomposition helps separate \emph{whether} UI should appear from \emph{how} it should be expressed.

\paragraph{Deterministic post-processing.}
After initial A2UI generation, we apply rule-based post-processing to correct frequent structural issues prior to validation. These fixes include enum normalization (e.g., icon names), data-binding type correction, field completion for partially specified components, and simple layout constraints needed by the renderer.

\paragraph{Component-targeted augmentation.}
To improve coverage of low-frequency components, we add 4{,}165 augmented samples, accounting for 29.2\% of the final training set. Augmentation is targeted rather than uniform: we primarily expand under-represented layout, interactive, and multimedia components such as rows, slider-like controls, icons, images, date/time inputs, modals, tabs, check boxes, video proxies, and audio proxies. This design increases structural diversity without overwhelming the corpus with synthetic negatives or unrealistic UI patterns.

\subsection{Validation and Repair}
\label{subsec:data-validation}

We validate all generated UI-turns with a four-level linting pipeline:
\emph{format} validation checks whether the response can be parsed as valid json structured output;
\emph{structure} validation checks required fields, component typing, and enum correctness;
\emph{data-binding} validation checks field/value compatibility and binding completeness;
and \emph{semantic} validation performs lightweight consistency checks between UI structure and intended interaction semantics.

Any generated sample that fails validation is retried with concise error feedback, for up to
three attempts per sample. After deterministic post-processing and lint validation, 91.3\%
of UI-turns pass on the first attempt. Error-feedback retry recovers an additional 7.6\%,
yielding a final renderability rate of 99.2\% over all UI-turns, with only 85 samples failing
after three attempts.

% \begin{table}[t]
% \centering
% \small
% \setlength{\tabcolsep}{6pt}
% \begin{tabular}{lcc}
% \toprule
% Validation outcome & Count & Share of UI-turns \\
% \midrule
% First-pass success & 9,320 & 91.3\% \\
% Fixed on 2nd attempt & 703 & 6.9\% \\
% Fixed on 3rd attempt & 75 & 0.7\% \\
% Failed after 3 attempts & 85 & 0.8\% \\
% \midrule
% Final renderable UI-turns & 10,125 / 10,210 & 99.2\% \\
% \bottomrule
% \end{tabular}
% \caption{Validation and repair statistics on UI-turns. First-pass success is measured after deterministic post-processing and lint validation.}
% \label{tab:validation-summary}
% \end{table}

\subsection{Dataset Statistics and Characteristics}
\label{subsec:data-stats}

\begin{figure}[t]
    \centering
    \begin{subfigure}[t]{0.48\columnwidth}
        \centering
        \includegraphics[width=\linewidth]{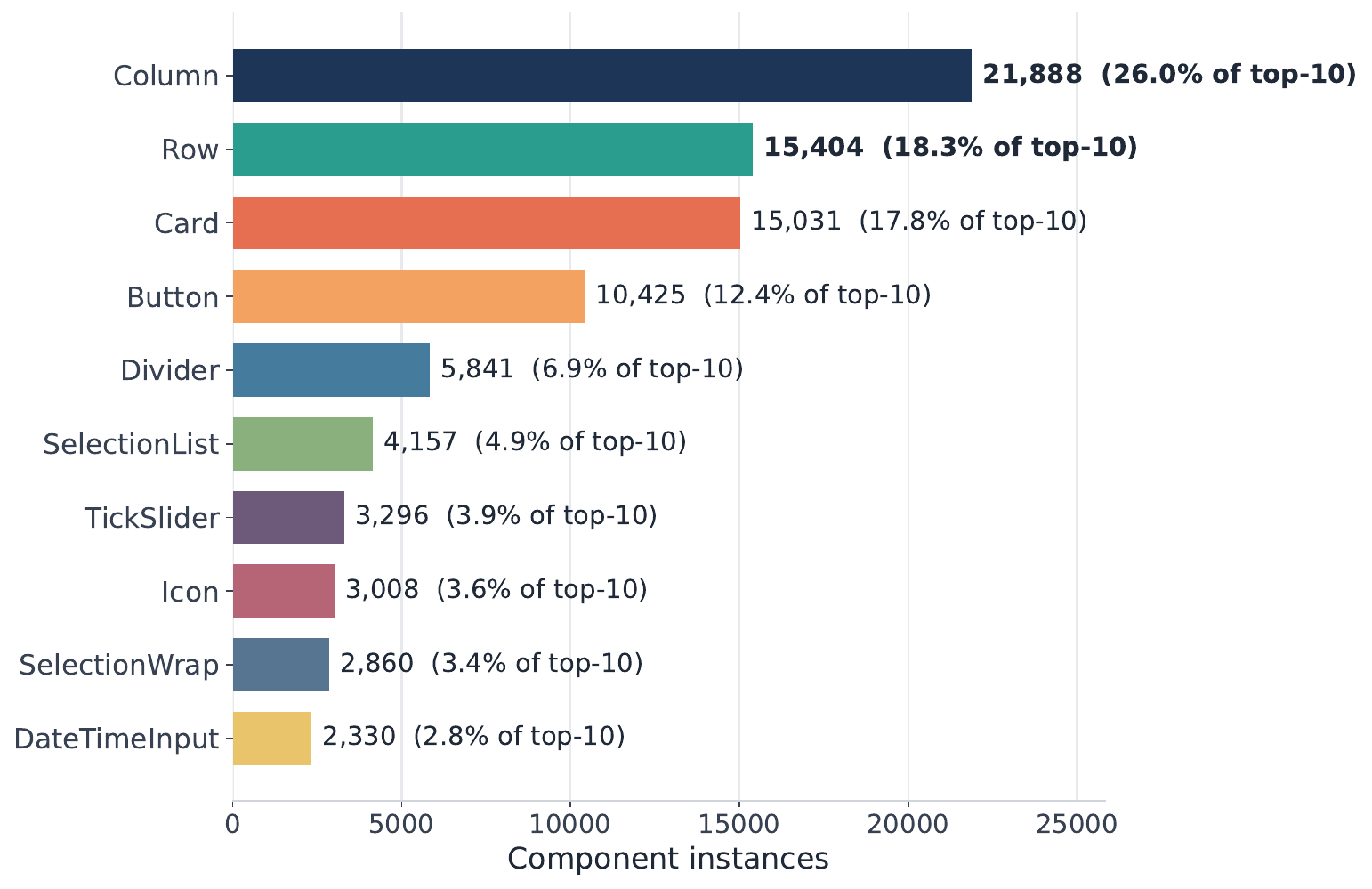}
        \caption{Top-10 component frequencies.}
        \label{fig:component-dist}
    \end{subfigure}\hfill
    \begin{subfigure}[t]{0.48\columnwidth}
        \centering
        \includegraphics[width=\linewidth]{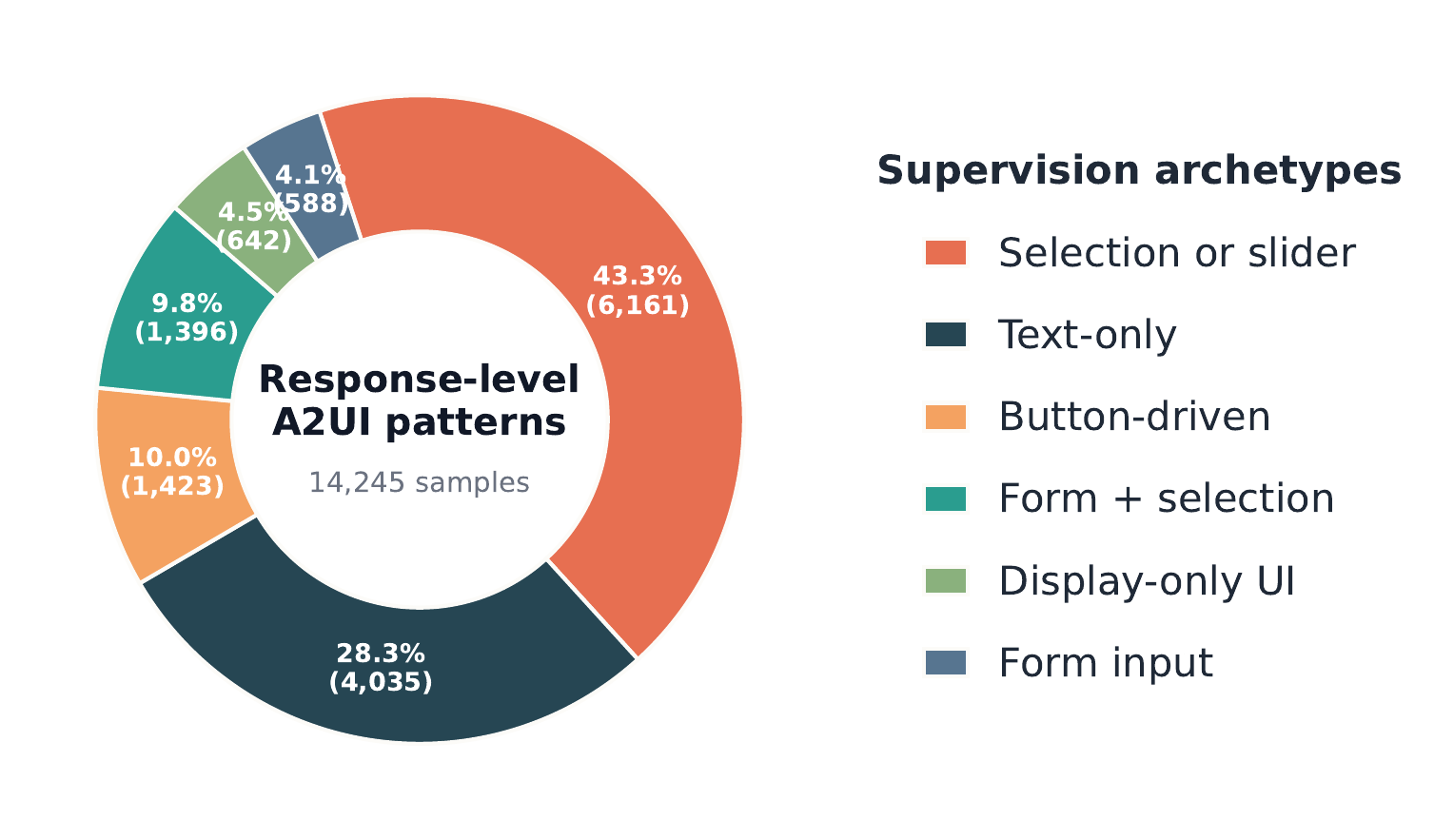}
        \caption{Response-level supervision archetypes.}
        \label{fig:data_coverage}
    \end{subfigure}
    \caption{Dataset statistics of the training corpus. (a) shows the top-10 component frequencies, and (b) shows the response-level supervision archetypes.}
    \label{fig:dataset_stats}
\end{figure}

As summarized in Table~\ref{tab:corpus_composition}, the final training set contains
14,245 assistant-turn samples, including 10,210 UI-turns and 4,035 text-only turns, yielding
an overall UI ratio of 71.7%.

% \begin{table}[t]
% \centering
% \small
% \setlength{\tabcolsep}{4.5pt}
% \begin{tabular}{lrrrrrr}
% \toprule
% Source & Total & Original & Aug. & UI & No-UI & UI ratio \\
% \midrule
% MultiWOZ & 5,424 & 3,673 & 1,751 & 4,361 & 1,063 & 80.4\% \\
% SGD & 4,757 & 3,692 & 1,065 & 3,761 & 996 & 79.1\% \\
% ESConv & 1,098 & 760 & 338 & 604 & 494 & 55.0\% \\
% AnnoMI & 2,966 & 1,955 & 1,011 & 1,484 & 1,482 & 50.0\% \\
% \midrule
% Total & 14,245 & 10,080 & 4,165 & 10,210 & 4,035 & 71.7\% \\
% \bottomrule
% \end{tabular}
% \caption{Per-source sample distribution. ``Original'' and ``Aug.'' denote samples originating from base dialogues and component-targeted augmentation, respectively.}
% \label{tab:data-distribution}
% \end{table}

This distribution reflects the interaction properties of the source corpora. Task-oriented datasets naturally have a high UI ratio, around 80\%, because many assistant turns correspond to structured information collection, result presentation, or confirmation. In contrast, ESConv and AnnoMI are intentionally closer to a balanced setting: in many supportive or counseling dialogues, pure text is more appropriate for conveying empathy, reflection, or self‑disclosure, and forcing UI into those turns would distort the original interaction style.

The text-only turns in our corpus are mostly natural rather than artificially synthesized. Of the 4{,}035 no-UI samples, 3{,}277 (81.2\%) are drawn directly from source dialogues, while only 758 (18.8\%) are introduced through augmentation.

At the component level, the corpus contains roughly 189k instances. Structural elements such as \texttt{Label}, \texttt{Column}, \texttt{Row}, and \texttt{Card} dominate the scaffolding of the UI, while interactive elements such as buttons, selection widgets, slider-like controls, and date/time inputs provide the supervision needed for user-facing interaction design. 
Figure~\ref{fig:component-dist} visualizes this component distribution and highlights the contribution of augmentation to long-tail coverage.

Figures~\ref{fig:data_coverage} provide a complementary view of corpus coverage. At the component level, the training data covers both common layout primitives and task-critical interactive widgets. Frequent structural components such as \texttt{Column}, \texttt{Row}, and \texttt{Card} provide supervision for compositional UI layout, while interactive components such as \texttt{Button}, \texttt{SelectionList}, \texttt{TickSlider}, \texttt{SelectionWrap}, and \texttt{DateTimeInput} expose the model to a broad range of executable interaction patterns. At the response level, the corpus is also diverse in supervision archetypes: beyond text-only turns, it contains substantial coverage of selection- or slider-based interaction, button-driven actions, mixed form-and-selection responses, display-oriented UI, and explicit form input. Together, these statistics suggest that the corpus does not merely teach surface syntax, but covers a meaningful subset of A2UI structures ranging from basic layout composition to interactive decision support and structured information collection.

% \section{Benchmark}
% \paragraph{Task taxonomy: atomic, depth, width}

% atomic: single step intent task

% depth emphasizes continuous UI decisions and state advancement in multi-round series connection.

% width emphasizes the combined expression of multiple parallel intents in a turn, or the coordination of multiple surfaces/components.

% \paragraph{Multi-level metrics}
% L1 represents the correctness at the protocol and format levels: whether it can be parsed, the legality of fields, and whether it can be rendered.

% L2 is the correctness at the task advancement level: whether the UI is consistent with the current semantics and whether it helps collect necessary information or advance to the next step.

% L3 refers to the quality at the user experience level: whether it is natural, low-friction, not excessive, not interrupted, and avoids unnecessary rounds and input costs.

% \paragraph{Evaluation protocol}

% \paragraph{SOTA model analysis}

\section{A2UI-Bench}
\label{sec:benchmark}

The A2UI corpus in Section~\ref{sec:data} is designed for large-scale supervision, whereas evaluation requires a different emphasis: controlled coverage, balanced task composition, and diagnostic scoring. We therefore construct \textsc{A2UI-Bench}, a dedicated benchmark derived from the same data construction framework but optimized for model assessment rather than training-scale diversity.

\textsc{A2UI-Bench} is designed to evaluate three aspects of Generative UI under A2UI. First, it should cover both common and difficult A2UI behaviors, including UI triggering, UI suppression, cross-turn consistency, and compositional organization. Second, it should distinguish low-level protocol correctness from higher-level functional quality and user experience. Third, it should support direct comparison between models through a fixed task composition and a shared evaluation protocol.

\subsection{Task Taxonomy and Benchmark Composition}
\label{subsec:bench-taxonomy}

We organize \textsc{A2UI-Bench} by \textbf{task structure}. This keeps the benchmark compact while directly targeting the structural capabilities that matter for Generative UI.

\paragraph{Atomic tasks.}
Atomic tasks are single-turn, single-intent evaluations. Given a dialogue context and the current user message, the model produces one assistant response that may include both text and an A2UI payload. These tasks measure the core turn-level ability to decide whether UI is needed and, if so, to generate a protocol-compliant and semantically appropriate interface.

\paragraph{Depth tasks.}
Depth tasks evaluate multi-turn consistency. Each task consists of a short episode of consecutive turns from the same dialogue. The evaluator rolls the interaction forward using the model's own previous output. This tests whether the model can maintain coherent state, update or replace previously rendered surfaces, and handle cross-turn dependencies.

\paragraph{Width tasks.}
Width tasks are single-turn but compositionally broader. Each task combines multiple information needs, often spanning more than one intent or service, into one user request. The model must organize a unified response that addresses several sub-goals without producing fragmented or cognitively heavy UI. These tasks emphasize structural organization and interaction planning.

Beyond task structure, the benchmark covers a diverse set of interaction intents, including structured parameter collection, information grounding, booking confirmation, decision support, self-assessment, action commitment, multi-part organization, tie-breaking, and explicit \texttt{no\_ui\_chat} cases. The last category is an important negative class: models are expected to avoid unnecessary UI when plain text is more appropriate.

The final benchmark contains 300 tasks. We intentionally include negative examples and structurally broader width tasks so that the benchmark does not collapse into a simple single-turn form-generation test.

\subsection{Metrics}
\label{subsec:metrics}

We evaluate A2UI generation from two complementary perspectives. The first is a language-side evaluation that scores protocol correctness, task construction quality, and interaction quality directly from the model output. The second is a visual-side evaluation that scores the rendered UI screenshot seen by the user.

\paragraph{Language-side metrics.}
The language-side evaluation contains three levels.

L1 measures protocol correctness through five automated dimensions: L1-1 JSON parse correctness, L1-2 schema compliance, L1-3 reference integrity, L1-4 required fields completeness, and L1-5 value format correctness. Its design follows a hierarchical reliability principle. If the output is unparsable, all five dimensions receive zero. If parsing succeeds but the validator detects structural errors, only L1-1 is credited and L1-2 through L1-5 are set to zero, since downstream checks are no longer reliable enough to support partial credit. Only when the output is error-free do we use warnings to deduct points from L1-2 through L1-5. As a result, L1 distinguishes between unparsable outputs, parseable but structurally invalid outputs, and structurally valid outputs that remain imperfect.

L2 measures task construction quality through five LLM-judged dimensions: L2-1 trigger appropriateness, L2-2 component--intent alignment, L2-3 text--UI grounding, L2-4 data model utilization, and L2-5 action completeness. L2-1 asks whether the model made the right decision to generate or suppress UI for the current turn. L2-2 evaluates whether the chosen widget type matches the interaction need. L2-3 requires every visible label, option, and value in the UI to be grounded in the accompanying text response. L2-4 evaluates whether \texttt{dataModelUpdate} is used appropriately to track dynamic state such as collected slots or user selections. L2-5 asks whether interactive components form a complete loop with a valid action and sufficient returned context for the next step.

L3 measures user experience quality through three LLM-judged dimensions: L3-1 value-addition over text, L3-2 conversational naturalness, and L3-3 cognitive load. L3-1 asks whether the UI meaningfully reduces user effort rather than merely repackaging text in visual form. L3-2 evaluates whether the transition from text to UI feels natural and contextually motivated. L3-3 measures whether the amount of information and interaction requested in a single turn remains manageable. Together, L2 and L3 distinguish functional correctness from interaction quality, allowing us to evaluate not only whether a model can produce valid A2UI, but also whether it produces useful and well-paced conversational interfaces.

This maps the language-side evaluation to the range $[0,1]$. We treat L1 as a reliability gate because structurally invalid outputs cannot be rendered or interacted with reliably; such failures also constrain downstream L2 and L3 scoring. This is consistent with the current benchmark aggregation used in the evaluation code and reports.

\paragraph{Visual-side metrics.}
The visual-side evaluation measures the quality of the rendered UI. For each visual target, we render the model-produced A2UI through the client renderer, capture a screenshot, and score it with a VLM judge. Atomic and width tasks yield one visual target per task, while depth tasks are expanded into step-level visual targets. Only targets with successful parsing, non-empty A2UI, and successful render check are included in visual evaluation.

The visual-side evaluation contains three dimensions:
V1 evaluates visual integrity, including readability, clipping, overflow, and obvious layout defects;
V2 evaluates task alignment, namely whether the visible UI matches the task and assistant response;
V3 evaluates action clarity, namely whether a user can clearly infer what to do next from the rendered interface.

\section{Experiment}
\label{sec:exp}
\subsection{Training Pipeline}
\label{subsec:training-pipeline}

\begin{table}[t]
\centering
\small
\setlength{\tabcolsep}{5pt}
\begin{tabular}{l|c|ccc|ccc|c}
\toprule
Model & Prompt & L1 & L2 & L3 & V1 & V2 & V3 & Avg. \\
\midrule
GPT-5.4 & w/ schema & 4.02 & \textbf{3.59} & 3.27 & 3.46 & 3.73 & 3.17 & 3.54 \\
Gemini-3.1-Pro & w/ schema & 4.25 & 3.20 & 2.96 & 3.53 & 3.55 & 3.04 & 3.42 \\
Qwen3-235B-A22B-Instruct & w/ schema & 4.00 & 2.87 & 2.76 & 3.32 & 3.32 & 2.95 & 3.20 \\ 
DeepSeek-V3.1 & w/ schema& 4.19 & 2.54 & 2.47 & 3.27 & 3.35 & 2.95 & 3.13 \\
Qwen3-30B-A3B-Instruct & w/ schema & 3.13 & 2.13 & 2.09 & 2.92 & 2.64 & 2.51 & 2.57 \\
GPT-4o-mini & w/ schema & 3.45 & 2.27 & 2.15 & 2.78 & 2.28 & 2.06 & 2.50 \\
\midrule
Macaron-A2UI-Grande & w/o schema & \textbf{4.67} & 3.22 & 2.91 & \textbf{3.95} & 3.74 & 3.47 & 3.66 \\
Macaron-A2UI-Venti & w/o schema & 4.47 &3.36 &\textbf{3.28} &\textbf{3.95} & \textbf{3.76} &\textbf{3.52} & \textbf{3.72} \\
\bottomrule
\end{tabular}
\caption{Full-schema prompted baselines and our schema-light model references. Models marked w/ schema receive the complete A2UI schema, while Macaron-A2UI models are evaluated without explicit schema hints. Higher is better on all metrics.}
\label{tab:full_prompt_results}
\end{table}

We train our A2UI-capable assistant with a parameter-efficient two-stage pipeline consisting of supervised fine-tuning (SFT) followed by group-relative policy optimization (GRPO). Both stages use LoRA adaptation, which updates a small set of low-rank parameters instead of fully tuning the backbone. We instantiate the same training recipe on Qwen3-30B-A3B-Instruct-2507 and Qwen3-235B-A22B-Instruct-2507, and further include GLM-5.1 as an additional backbone and keep the output protocol fixed across all stages. In both cases, the model is trained to produce a unified assistant response that combines natural language and structured A2UI actions.

The role of the two stages is complementary. SFT teaches the model the basic response format, including how to jointly produce fluent text and protocol-compliant UI actions. GRPO then refines this behavior under an interaction-oriented reward.

\paragraph{Supervised fine-tuning.}
Our SFT data are organized as chat-style prompt-response pairs. Each sample contains a prompted conversation consisting of the system instruction and dialogue history, and a target assistant response. The target always follows the same JSON schema,
\(
\{\texttt{"text\_response"}: \ldots, \texttt{"a2ui"}: [\ldots]\},
\)
where \texttt{text\_response} is the response in natural-language and \texttt{a2ui} is a list of actions of the structured protocol. During training, we concatenate the prompted conversation and the target completion into one conversational sequence and compute loss only on the final assistant turn. Given context $x$ and target response $y=(y_1,\dots,y_T)$, the SFT objective is the standard autoregressive negative log-likelihood
\[
\mathcal{L}_{\mathrm{SFT}}
=
-\sum_{t=1}^{T}\log p_{\theta}(y_t \mid x, y_{<t}).
\]
This stage directly teaches the model to jointly generate text and executable UI actions, rather than treating interface generation as a separate post-processing step.

\paragraph{GRPO optimization.}
Starting from the SFT model, we further optimize the same chat-style interface with GRPO. For each prompt $x_i$, we sample a group of $G$ candidate responses $\{y_{i,1}, \dots, y_{i,G}\}$ from the current policy and score each candidate with our A2UI reward function. We then compute a group-relative advantage by centering rewards within each sampled group
\[
A_{i,j}
=
R_{i,j}
-
\frac{1}{G}\sum_{k=1}^{G} R_{i,k}.
\]
This advantage is assigned to all tokens in the corresponding response, and the optimization favors candidates that outperform their within-group alternatives under the same prompt. The resulting objective is
\[
\mathcal{L}_{\mathrm{GRPO}}
=
-\sum_i \sum_{j=1}^{G}\sum_{t=1}^{|y_{i,j}|}
A_{i,j}
\log p_{\theta}(y_{i,j,t} \mid x_i, y_{i,j,<t}).
\]
Compared with SFT, this stage is better suited to properties that are difficult to supervise with a single gold target, such as UI triggering, action completeness, and interaction quality across structurally different but valid responses.

\paragraph{Reward design.}
Our GRPO reward is designed around executable A2UI quality rather than text quality alone. We first apply hard structural gates: malformed JSON, missing required A2UI output when UI is expected, protocol-level validation failure, or render-critical structural errors all receive zero reward. For responses that pass these checks, the final reward is a weighted combination of structural quality, task-construction quality, and user-level utility:
\[
R
=
\mathbf{1}[\mathrm{pass}]
\cdot
\left(
\lambda_1 S_{\mathrm{L1}}
+
\lambda_2 S_{\mathrm{L2}}
+
\lambda_3 S_{\mathrm{L3}}
\right).
\]
Here, $S_{\mathrm{L1}}$ measures low-level correctness, including JSON validity, schema compliance, reference integrity, required-field completeness, value-format correctness, and renderability. $S_{\mathrm{L2}}$ measures task-construction quality, including trigger appropriateness, component--intent alignment, text--UI grounding, data-model utilization, and action completeness. $S_{\mathrm{L3}}$ measures user-level utility, including value-add over plain text, conversational naturalness, and cognitive load. For no-UI cases, we use a simplified reward that explicitly favors text-only responses when interaction is unnecessary. Overall, this design aligns reinforcement learning with the central goal of our setting: producing responses that are simultaneously executable, appropriate, and genuinely helpful in interaction.

\begin{figure*}[t]
    \centering
    \includegraphics[width=\textwidth]{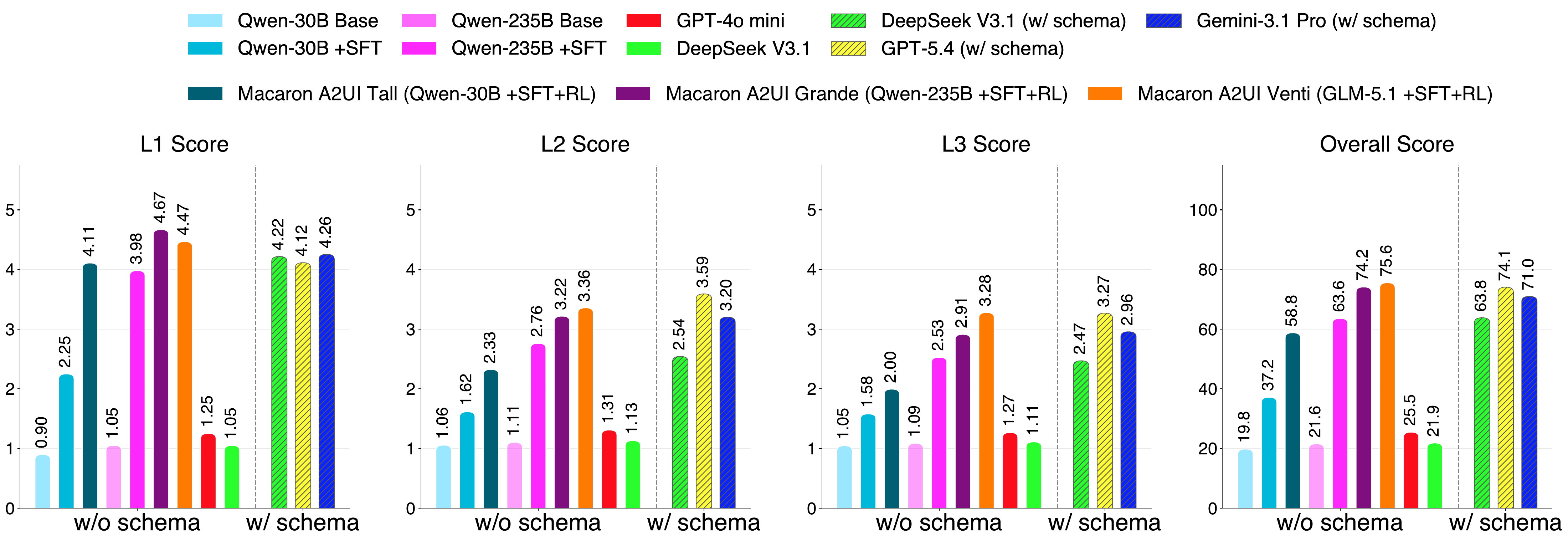}
    \caption{Training-pipeline ablation under the w/o schema prompt regime and comparison to full-prompt upper bounds. Solid bars on the left of each panel show minimal-prompt results for untuned, SFT, and SFT+RL models, together with untuned frontier references. Hatched bars on the right of the dashed separator show full-prompt upper bounds, where models receive the complete A2UI schema and protocol specification. Scores are reported on A2UI-Bench for L1, L2, L3, and the overall language-side score.}
    \label{fig:minimal_vs_full_upperbound}
\end{figure*}

\subsection{Main Results}
\label{subsec:main-results}
\begin{wrapfigure}{r}{0.52\linewidth}
    \centering
    \vspace{-14pt}
    \includegraphics[width=\linewidth]{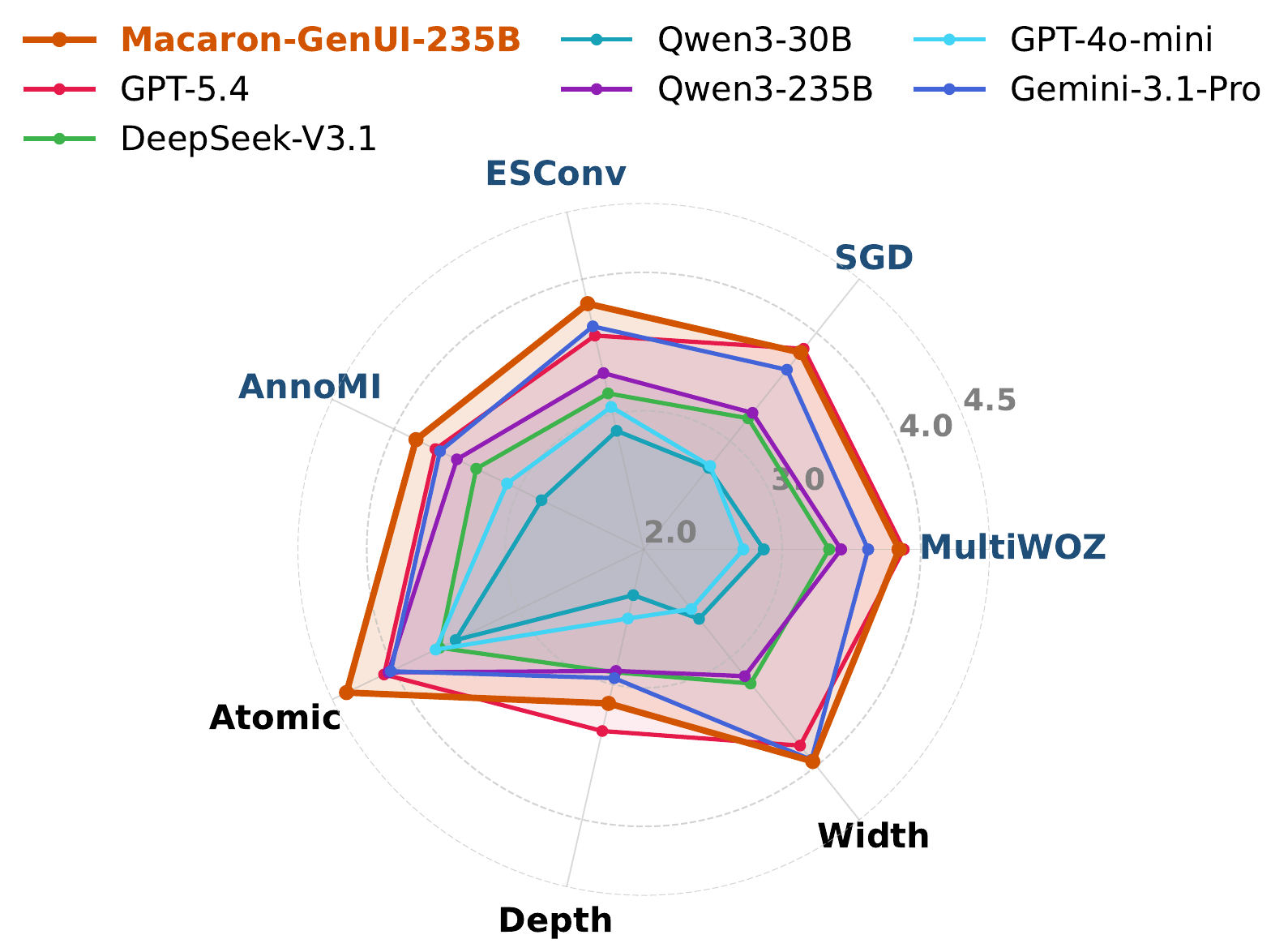}
    \vspace{-10pt}
    \caption{Per-dataset and per-task breakdown of model performance on A2UI-Bench.}
    \label{fig:radar}
    \vspace{-12pt}
\end{wrapfigure}

Figure~\ref{fig:minimal_vs_full_upperbound} summarizes our main results under two evaluation regimes. The primary regime is the prompt w/o schema setting, where models receive only lightweight protocol instructions and must rely on internalized A2UI competence. This is the main setting for evaluating our training pipeline. We additionally report results w/ schema as prompt-heavy upper bounds, where models are given the complete A2UI schema and protocol specification.

The results show that supervised fine-tuning is highly effective on both model scales. For Qwen-30B, SFT improves the overall score from 19.8 to 37.2, with especially large gains on L1 (0.90 to 2.25). RL further increases the overall score to 58.8 and pushes L1 to 4.11, while also improving L2 and L3. The same trend holds even more strongly with the scaling of model size. Qwen-235B improves from 21.6 at base to 63.6 after SFT, and then further reaches 74.2 after RL. Beyond the Qwen backbones, Macaron-A2UI-Venti, trained from GLM-5.1 with the same SFT+RL recipe, obtains the strongest language-side result in Figure~\ref{fig:main_results}, with an overall score of 75.6.

A second observation is that out-of-the-box frontier models remain weak without schema hints in prompts. GPT-4o mini, DeepSeek-V3.1, and GPT-5.4 obtain overall scores of 25.5, 21.9, and 23.9, respectively, far below the trained open models. Their L1, L2, and L3 scores are all low, suggesting that lightweight instructions are insufficient for untuned general-purpose models to acquire stable A2UI competence. In contrast, our tuned models improve substantially without requiring the full protocol specification at inference time.

The prompt w/ schema settings, serving as upper bounds (Table~\ref{tab:full_prompt_results}) provide a complementary picture. When given the complete schema, frontier models become much stronger: DeepSeek-V3.1 reaches 63.8 overall, Gemini-3.1 Pro reaches 71.0, and GPT-5.4 reaches 74.1. This confirms that prompt-heavy schema injection can unlock strong A2UI performance for frontier models. 
At the same time, our schema-light trained models are competitive with or stronger than these prompt-heavy references. Macaron-A2UI-Grande reaches 74.2, slightly surpassing GPT-5.4 with the full schema, while Macaron-A2UI-Venti further improves the overall score to 75.6.

Figure~\ref{fig:radar} shows a more structured picture than a simple average improvement. Since the evaluation set is balanced across the four source datasets and three task formats, the observed results are not driven by source skew. Macaron-A2UI-235B is the strongest model overall (3.83), outperforming GPT-5.4 (3.75) and substantially improving over the untuned Qwen3-235B backbone (3.37). Scores on MultiWOZ, SGD, ESConv, and AnnoMI all fall in a narrow range (3.82–3.84), indicating strong cross-domain robustness. Macaron-A2UI-235B is the best model in atomic tasks (4.38) and width tasks (3.96), while remaining second in depth (3.14, behind GPT-5.4’s 3.34). This suggests that RL mainly strengthens the model’s ability to translate dialogue intent into concise, well-structured, and interaction-ready UI decisions. On the dataset level, Macaron-A2UI-235B is the top model on ESConv and AnnoMI, and relative to Qwen3-235B it improves width consistently across all four datasets, with especially large gains on SGD and MultiWOZ.

\begin{figure}[ht]
    \centering
    \begin{subfigure}[t]{0.24\columnwidth}
        \centering
        \includegraphics[width=\linewidth]{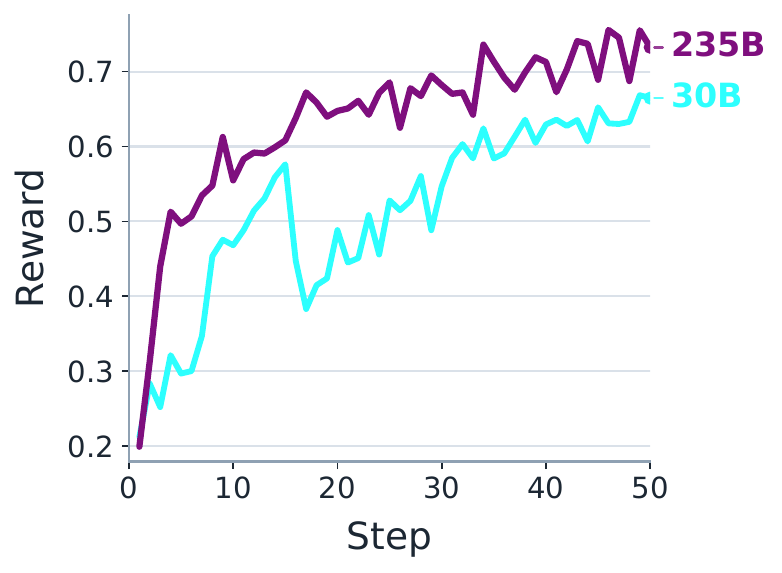}
        \caption{Total reward}
        \label{fig:total_reward}
    \end{subfigure}\hfill
    \begin{subfigure}[t]{0.24\columnwidth}
        \centering
        \includegraphics[width=\linewidth]{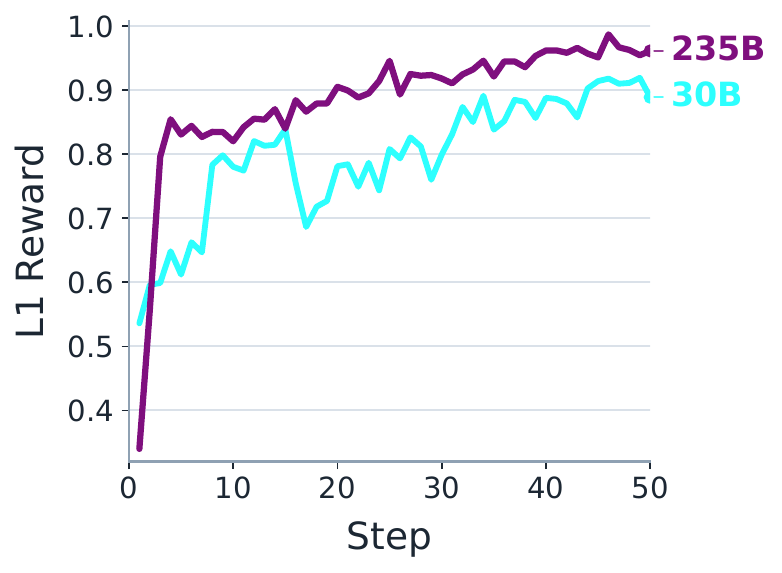}
        \caption{L1 reward}
        \label{fig:l1_reward}
    \end{subfigure}\hfill
    \begin{subfigure}[t]{0.24\columnwidth}
        \centering
        \includegraphics[width=\linewidth]{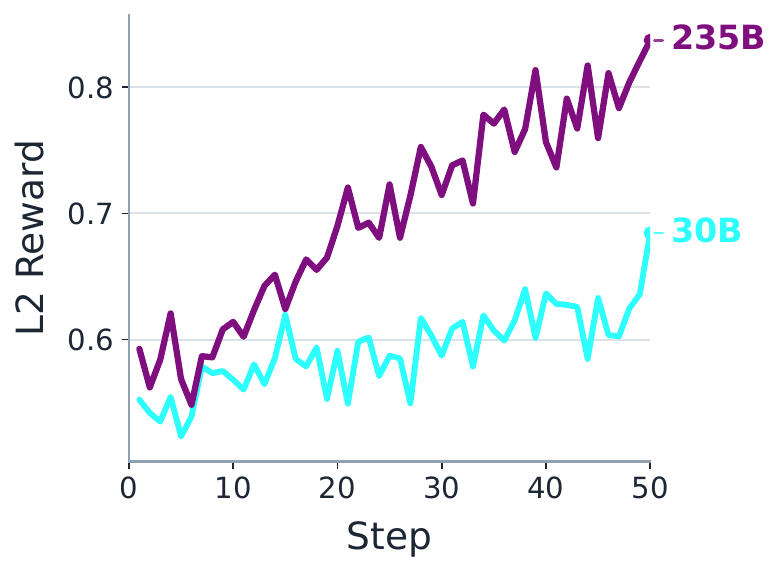}
        \caption{L2 reward}
        \label{fig:l2_reward}
    \end{subfigure}\hfill
    \begin{subfigure}[t]{0.24\columnwidth}
        \centering
        \includegraphics[width=\linewidth]{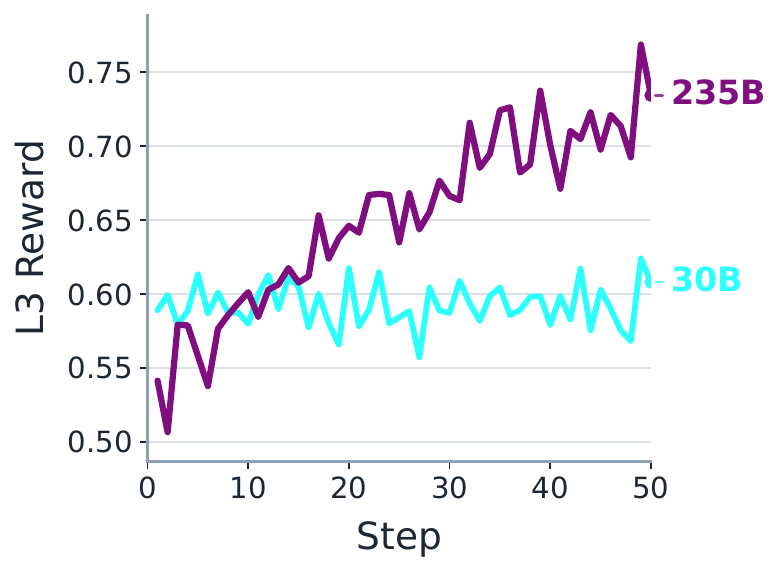}
        \caption{L3 reward}
        \label{fig:l3_reward}
    \end{subfigure}\hfill
    \caption{Reward trajectories during GRPO training. We track the total reward together with its L1, L2, and L3 components for Macaron-A2UI-30B and Macaron-A2UI-235B over training steps. The L1 reward rises first and most rapidly, while L2 and L3 improve more gradually. Higher is better.}
    \label{fig:reward}
\end{figure}

\paragraph{RL training dynamics.}
Figure~\ref{fig:reward} shows the reward trajectories during GRPO training. The optimization dynamics are highly consistent with the design of our reward. Across both model scales, the L1 reward increases first and most rapidly, indicating that protocol correctness and structural executability are the easiest properties to improve under reinforcement learning. This is expected, since malformed outputs, schema violations, and render-critical errors are directly penalized and provide a relatively stable learning signal. In contrast, improvements in higher-level interaction quality occur more gradually. For the 235B model, L2 and L3 rewards increase steadily throughout training, suggesting that once protocol competence is stabilized, the larger model can continue refining task construction quality and user-facing interaction quality in a relatively balanced way. The 30B model exhibits a different pattern: while its L2 reward still improves over training, its L3 reward remains much flatter and is clearly the hardest component to optimize.

\section{Conclusion}

In this paper, we study A2UI-based Generative UI as a unified problem in which the assistant produces both natural language and executable UI actions under a fixed declarative protocol. To support this setting, we construct an A2UI-grounded corpus from heterogeneous dialogue sources, build A2UI-Bench for controlled evaluation, and develop a two-stage training pipeline with schema-light SFT and reward-driven RL. Experiments on both 30B and 235B models show that this training recipe improves protocol correctness, interaction quality, and user experience by a large margin. The best 235B model performs strongly in the minimal-prompt setting and slightly surpasses the strongest full-prompt frontier baseline in overall score. These results show that Generative UI capability does not have to depend on heavy schema prompting at inference time, and can instead be learned and internalized through training.

This project is a key step toward bringing Generative UI into real production environments, but it still has clear limitations. The underlying A2UI protocol is still evolving and currently remains at version 0.8. Model capability is also a bottleneck, especially for complex multi-turn interaction and user experience. In addition, latency can become an issue when the model must generate, validate, and render structured UI in real time. In future work, we will continue to explore more general, flexible, and token-efficient ways to build Generative UI systems. We release our models, benchmark, and evaluation protocol to support broader research and development in this direction.

\bibliography{colm2026_conference}
\bibliographystyle{colm2026_conference}

\appendix
\section{Appendix}
\subsection{A2UI Rendering Implementation}
\label{appendix:rendering}

To evaluate the visual quality of model-generated A2UI outputs, we require a faithful rendering implementation that transforms the raw JSON message stream into a fully interactive widget tree. We implement this renderer as a Flutter Web application, which serves both as the production chat client and as the headless rendering backend for automated visual evaluation. This subsection describes the renderer architecture, the message protocol it consumes, and the full component catalog it supports.

\paragraph{Message Protocol and Rendering Pipeline.}
An A2UI output from the model is a JSON array of \emph{messages}, each carrying exactly one of four action types. A \texttt{beginRendering} message declares a new UI surface by specifying a \texttt{surfaceId} and the root component tree to display. A \texttt{surfaceUpdate} message patches an existing surface by replacing, inserting, or removing subtrees at specified paths within the component hierarchy. A \texttt{dataModelUpdate} message writes key-value pairs into a reactive data model that components can bind to via JSON pointer paths, enabling dynamic state such as pre-filled form values, selection defaults, and conditional visibility. A \texttt{deleteSurface} message removes a previously rendered surface entirely. The renderer processes these messages sequentially through an \texttt{A2uiMessageProcessor}, which maintains the current surface state and incrementally reconciles the Flutter widget tree after each message. This incremental design means that the renderer does not simply re-render from scratch on each update; instead, it applies structural diffs, which is important both for production responsiveness and for faithfully reflecting the multi-message outputs that models produce.

\paragraph{Component Catalog.}
The renderer resolves component types through a pluggable \emph{catalog} system. Each catalog entry (a \texttt{CatalogItem}) maps a component type name and an optional \texttt{extensionType} to a Flutter widget builder, a JSON Schema describing the component's data contract, and metadata about whether the component accepts children. Our evaluation uses the Macaron design catalog, which registers 23 component types organized into four functional categories. Table~\ref{tab:component-catalog} provides the complete listing. \emph{Interactive} components accept user input and can dispatch actions back to the conversational agent (e.g., a selection event, a form submission, or a button tap), which is the core mechanism through which A2UI enables structured user responses beyond free-text typing. \emph{Display} components present information without accepting input, providing the visual scaffolding (text, images, progress indicators) that contextualizes the interactive elements. \emph{Layout} components control spatial arrangement of their children without contributing semantic content themselves.

\begin{table}[h]
\centering

\small
\begin{tabular}{llp{7.2cm}}
\toprule
\textbf{Component} & \textbf{Role} & \textbf{Description} \\
\midrule
\multicolumn{3}{l}{\textit{Selection \& Input}} \\
SelectionList & Interactive & Vertical list with check indicators; supports multi-select with min/max constraints \\
SelectionGrid & Interactive & Adaptive 1--3 column grid with overlay check indicators \\
SelectionWrap & Interactive & Horizontal flow wrap with pill-style chips for compact multi-select \\
OrderedSelectionList & Interactive & Vertical list showing pick order as numbered badges \\
ActionSelectionList & Interactive & Single-select list that locks and dispatches an action on first tap \\
DropdownSelection & Interactive & Button-like trigger with anchored dropdown menu for single-select \\
DateTimeInput & Interactive & Date and/or time picker trigger with localized display formatting \\
PasswordKeypad & Interactive & Six-digit numeric keypad with auto-submit on completion \\
\midrule
\multicolumn{3}{l}{\textit{Scalar \& Continuous Input}} \\
TickSlider & Interactive & Discrete slider snapping to evenly spaced tick marks \\
\midrule
\multicolumn{3}{l}{\textit{Action}} \\
Button & Interactive & Three visual styles (primary, secondary, plain) with action dispatch \\
\midrule
\multicolumn{3}{l}{\textit{Navigation \& Overlay}} \\
Tabs & Interactive & Capsule-track tab bar with animated thumb and swipe-enabled pages \\
Carousel & Interactive & Horizontal page view with peek-through adjacent pages \\
FullScreenModal & Interactive & Entry-point child triggers a full-screen overlay for detail content \\
\midrule
\multicolumn{3}{l}{\textit{Text \& Rich Content}} \\
Label & Display & Styled text with 3 color variants $\times$ 9 typographic sizes (27 combinations) \\
MarkdownView & Display & Renders markdown (headings, emphasis, lists, code blocks, blockquotes) \\
\midrule
\multicolumn{3}{l}{\textit{Data Visualization \& Media}} \\
LinearProgress & Display & Horizontal progress bar with animated fill \\
CircularProgress & Display & Circular arc indicator with optional center icon and value label \\
Image & Display & Network or asset image in four size presets (40--full width) \\
Icon & Display & SVG icon from a custom registry with line and filled styles \\
OrderedDisplayList & Display & Static vertical list with fixed-order numbered badges \\
\midrule
\multicolumn{3}{l}{\textit{Layout \& Structure}} \\
Card & Layout & Rounded container with warm background and vertical child spacing \\
Column & Layout & Vertical stack with configurable spacing, distribution, and alignment \\
Row & Layout & Horizontal stack with spacing and expansion semantics \\
Divider & Layout & Dashed line separator (horizontal or vertical) \\
\bottomrule
\end{tabular}

\caption{A2UI component catalog used in our evaluation. Components are grouped by functional role. ``Interactive'' components accept user input and can dispatch actions; ``Display'' components present information read-only; ``Layout'' components arrange children spatially.}
\label{tab:component-catalog}
\end{table}

\paragraph{Preview Stage and Render Page.}
For automated evaluation, the Flutter application exposes a dedicated \texttt{/render} route that accepts the full A2UI message array as a URL query parameter. The route parses the JSON payload, extracts the \texttt{surfaceId} from the first message that contains one (falling back to a default identifier), and feeds the messages into the same \texttt{A2uiRenderer} and \texttt{A2uiMessageProcessor} used in the production chat interface. The rendered widget tree is displayed inside a \texttt{PreviewStage} container---a width-constrained, rounded-corner card with configurable dimensions (default 420\,px wide, up to 1600\,px tall) and inner padding (24\,px). The stage applies a subtle drop shadow and clips overflow to its rounded boundary, simulating the visual context of a chat bubble. Once the Flutter layout engine completes its build and layout passes, the page signals readiness by setting a DOM attribute (\texttt{data-render-status="ready"}) on the document body. This attribute serves as the synchronization signal for the external screenshot tool. The use of a real Flutter widget tree, rather than a mock renderer or static HTML approximation, ensures that the screenshots faithfully reflect the exact visual output that end users would see, including all spacing, overflow, clipping, and interactive-state rendering behaviors that are impossible to predict from the JSON alone.

\subsection{VLM-Based Visual Evaluation}
\label{appendix:visual-eval}

Beyond the protocol-level (L1) and LLM-judge-based (L2/L3) evaluation described in the main text, we introduce an auxiliary visual evaluation layer that assesses the \emph{actual rendered appearance} of model-generated A2UI outputs. The motivation is straightforward: a model output can be structurally valid JSON that satisfies all schema constraints, yet still produce a visually broken or misleading user interface when rendered by a real UI framework. To close this gap, we implement a fully automated pipeline that captures a screenshot of the rendered output (using the renderer described in \S\ref{appendix:rendering}) and submits it alongside dialogue context to a vision-language model (VLM) for structured quality assessment.

\paragraph{Render Check Gate.}
Before entering the visual evaluation pipeline, each sample must pass a lightweight \emph{render check} that detects seven categories of render-critical defects not covered by the schema validator. These rules were derived from failure patterns observed during iterative development of the A2UI renderer and codified as deterministic checks over the raw JSON message array. Specifically, the render check verifies: (1)~selection-type components that reference a data model path must explicitly provide a \texttt{literalArray} field, since the renderer cannot resolve dynamic bindings at screenshot time; (2)~\texttt{TickSlider} widgets must not appear as direct children of \texttt{Row} layouts, which causes an unbounded-width crash in Flutter's flex algorithm; (3)~a single assistant response must contain at most one \texttt{surfaceId}, as the renderer instantiates one surface per response and multiple IDs produce undefined behavior; (4)~\texttt{Button.action.context} must be a JSON array rather than a dictionary, matching the renderer's deserialization expectation; (5)~selection item values must be plain strings rather than wrapped objects such as \texttt{\{literalString: ...\}}; (6)~\texttt{DateTimeInput} temporal bounds (\texttt{firstDate}, \texttt{lastDate}) must be ISO-8601 strings and boolean flags (\texttt{enableDate}, \texttt{enableTime}) must be actual booleans; and (7)~every \texttt{dataModelUpdate} message must include a \texttt{path} field. Samples that fail any of these checks are excluded from the visual evaluation and additionally receive a floor penalty of 1.0 on all L2 and L3 dimensions in the main evaluation, since a render-breaking output cannot meaningfully serve the user regardless of its semantic quality.

\paragraph{Screenshot Capture and Cropping.}
Samples that pass the render check are rendered by the Flutter Web application described in \S\ref{appendix:rendering}. We use Playwright with a Chromium backend to navigate to the render URL, wait for the \texttt{data-render-status="ready"} signal, and capture a full-page screenshot. Because full-page mode can produce excessive whitespace for short UIs while preserving scrollable content for long ones, we apply an automatic cropping step: the script identifies the page background color from corner pixel samples, scans inward from all four edges to find the bounding box of non-background content, and crops to that region with a small margin to retain card borders and drop shadows. For multi-turn (depth) evaluation tasks, the visual pipeline does not produce a single composite screenshot; instead, each episode step is treated as an independent visual evaluation target with its own per-step A2UI messages, dialogue context, and user message. Target identifiers follow the convention \texttt{\{task\_id\}\_\_step\{XX\}} to maintain traceability, ensuring that intermediate-turn visual quality issues are not masked by a final-turn screenshot that may happen to look correct.

\paragraph{VLM Judge.}
Each cropped screenshot is submitted to a VLM judge along with the task description, scenario definition, dialogue context, current user message, and the assistant's text response. The judge is instructed to evaluate strictly based on what is \emph{visibly shown} in the screenshot, without assuming that off-screen or hidden content is acceptable. The evaluation uses three dimensions, each scored on a 1--5 integer scale. \textbf{V1 (Visual Integrity)} assesses whether the rendered UI is legible, well-bounded, and visually stable, penalizing clipping, overflow, text touching edges, cramped layouts, broken widgets, low contrast, visible error text, and large empty regions that make the interface feel poorly composed. \textbf{V2 (Task Alignment)} checks whether the visible labels, options, values, and information hierarchy match the assistant's text response and the current task context, penalizing mismatched content, missing key information, misleading wording, and UIs that appear tidy but do not actually serve the user's current need. \textbf{V3 (Action Clarity)} evaluates whether the user can determine the next action from the screenshot alone, penalizing hidden or weak primary controls, confusing affordances, missing selection state, overcrowded interactive elements, and any layout that would leave the user uncertain about what to do next. The judge must return a structured JSON object with a per-dimension score and short reason, a list of detected issues, and a one-sentence overall note. Responses that omit any dimension, contain non-numeric scores, or report scores outside the 1--5 range are rejected and retried up to three times before being marked as judge failures.

\paragraph{Relationship to L1--L3 Evaluation.}
The visual evaluation layer is complementary to, rather than redundant with, the three-level main evaluation. L1 verifies protocol and schema correctness through deterministic rules; L2 and L3 assess task construction quality and interaction experience quality through LLM judges that read the raw A2UI JSON. The visual layer instead evaluates the \emph{end-to-end rendered result} as a user would perceive it, capturing rendering-specific defects (layout overflow, widget crashes, cropped text) that are invisible at the JSON level. In particular, two outputs with identical L1/L2/L3 scores can produce very different visual outcomes if one triggers a subtle renderer edge case. The visual evaluation is therefore best understood as a ``last mile'' quality check on the actual user-facing artifact.

\section{Experiments}
\subsection{Training Details}
\begin{table*}[t]
  \centering
  \small
  \setlength{\tabcolsep}{8pt}
  \begin{tabular}{lll}
  \toprule
  \textbf{Parameter} & \textbf{SFT} & \textbf{GRPO} \\
  \midrule
  Training objective & Supervised fine-tuning & Group Relative Policy Optimization \\
  LoRA rank & 16 & 16 \\
  Batch size & 32 & 32 \\
  Learning rate & $1\times10^{-4}$ & $3\times10^{-5}$ \\
  Training length & 1 epoch & 50 steps \\
  Sequence / generation length & max length $=4096$ & max generation tokens $=2200$ \\
  Optimizer $\beta_1$ & 0.9 & / \\
  Optimizer $\beta_2$ & 0.95 & / \\
  Optimizer $\epsilon$ & $1\times10^{-8}$ & / \\
  Random seed & 42 & shuffle seed $=0$ \\
  Group size & / & 8 \\
  Sampling temperature & / & 1.0 \\
  Sampling concurrency & / & 8 \\
  Judge model & / & \texttt{openai/gpt-5.1} \\
  Judge concurrency & / & 32 \\
  Reward weight ($L1$) & / & 0.2 \\
  Reward weight ($L2$) & / & 0.4 \\
  Reward weight ($L3$) & / & 0.4 \\
  \bottomrule
  \end{tabular}
  \caption{Main model-agnostic hyperparameters used in SFT and GRPO. We omit dataset paths, model names, logging options, preview-saving settings,
  prompt-mode switches, and other engineering details that do not materially affect the training rule itself.}
  \label{tab:appendix_training_hparams}
  \end{table*}

Our training pipeline contains two stages: supervised fine-tuning (SFT) followed by GRPO.
  For clarity and reproducibility, we report only the model-agnostic hyperparameters that directly affect optimization, sampling, or reward
  composition, while omitting engineering-specific options such as logging, checkpoint naming, preview dumping, prompt-format switches, and task
  filtering flags.
  For SFT, we used LoRA adaptation with standard Adam-style optimization.
  For GRPO, we optimized grouped rollouts with judge-based rewards, where the final reward combined structural quality and higher-level judge scores
  using fixed weights.
  Table~\ref{tab:appendix_training_hparams} summarizes the main settings.
\section{A2UI Prompts}
\label{app:a2ui-prompts}

This appendix documents representative prompt templates used in the A2UI pipeline, including the minimal/full system prompts, the L2/L3 LLM-judge prompts, and the V1--V3 VLM-judge prompt. Runtime placeholders are kept as-is. Long schema blocks are abbreviated for space; the released evaluation package will include the complete prompt files and schemas used for reproduction.

\subsection{System Prompts}
\label{app:a2ui-system-prompts}

\paragraph{System prompt w/o A2UI schema}
\begin{minted}[escapeinside=||, fontsize=\small, breaklines, breaksymbol=, breaksymbolleft=, breaksymbolright=, breakanywhere, bgcolor=gray!10]{markdown}
You are a conversational AI assistant. Reply with natural language text and optional A2UI messages.

Always output valid JSON:
{"text_response": "...", "a2ui": [...]}

# A2UI Protocol

Allowed message types (each message must contain exactly one action key):
- beginRendering: create/start rendering a surface
- surfaceUpdate: define/update the component tree on a surface
- dataModelUpdate: write/update data model values
- deleteSurface: remove a surface

Component item format:
{"id": "x", "component": {"TypeName": {...}}}

Use ONLY these component names:
- Button: Clickable action trigger. Usually include `action.name` and connect visible content via `child`.
- Card: Single-card container shell. Use `child` as the main root content of this card.
- Column: Vertical layout container. Arrange child component IDs in `children`.
- DateTimeInput: Date/time picker input. Main binding field is `value`.
- Divider: Visual separator line. Use `axis` to choose horizontal or vertical.
- FullScreenModal: Full-screen modal container. Wire `entryPointChild` and `contentChild`.
- Icon: Icon display component. Field `name` is `{literalString: "icon-name"}`, `style` is `line` or `filled`.
  Valid icon names (use ONLY these): star, home, search, time, like, dislike, thumbs-up, thumbs-down,
  success, tips, fire, lightning, protection, alarm, alarm-clock, calendar-thirty, stopwatch, hourglass-null,
  arrow-left, arrow-right, arrow-circle-up, arrow-circle-down, arrow-circle-left, arrow-circle-right,
  book, book-one, book-open, notes, copy, link, share, share-two, rss, history, refresh,
  phone-telephone, mail-open, camera, pic-one, local-two, shopping-bag-one,
  knife-fork, chef-hat-one, cook, bowl, pot, platte, goblet, tea-drink, avocado-one, cheese, refrigerator,
  birthday-cake, leaves-two, sleep, abdominal, afferent,
  smiling-face-with-squinting-eyes, grinning-face-with-tightly-closed-eyes-open-mouth,
  anguished-face, disappointed-face, emotion-unhappy,
  more, more-one, hamburger-button, all-application, setting-three, equalizer, application-effect,
  preview-open, preview-close-one, left-c, right-c.
- Image: Image display component. Main field is `url`.
- Label: Plain text display component. Main field is `text`, optional `variant`.
- MarkdownView: Rich text/markdown display. Main field is `text`.
- PasswordKeypad: Secure keypad input. Provide `value.path` and submission `action`.
- Row: Horizontal layout container. Arrange child component IDs in `children`.
- SelectionList: Option selection list. Core fields are `selection` and `items`.
- Tabs: Tabbed content switcher. Define tab entries in `tabItems`.
- TickSlider: Discrete slider input. Core fields are `value` and `max`.
\end{minted}

\paragraph{System prompt w/ A2UI schema}
\begin{minted}[escapeinside=||, fontsize=\small, breaklines, breaksymbol=, breaksymbolleft=, breaksymbolright=, breakanywhere, bgcolor=gray!10]{markdown}
You must strictly follow schema field names; do not use onClick/onPress/content or other undefined names. Available components (only these 15): Button, Card, Column, DateTimeInput, Divider, FullScreenModal, Icon, Image, Label, MarkdownView, Row, SelectionList, SelectionWrap, Tabs, TickSlider.

## A2UI Message Protocol (CRITICAL)

The `a2ui` array is a list of **messages**. Each element must be a message object with **exactly ONE** action key (do NOT put raw component items like {"id": "...", "component": {...}} at the top level).

- `beginRendering`: {surfaceId: string, root: string} --- create surface; root is the ID of the top-level component.
- `surfaceUpdate`: {surfaceId: string, components: array} --- define the component tree. Each item in components: {"id": "unique_id", "component": {"TypeName": {...props}}}.
- `dataModelUpdate`: {surfaceId: string, path: "/", contents: array} --- write data. Each entry: {key: "name", valueString: "text"} or valueNumber / valueBoolean.
- `deleteSurface`: {surfaceId: string} --- remove surface.

Correct pattern: use ONE surfaceId; include beginRendering (root pointing to top component), then surfaceUpdate (with components array), then dataModelUpdate if needed. Components go **inside** surfaceUpdate.components, not directly in a2ui.

## Key Rules
1. Every surfaceUpdate must have a matching beginRendering with root pointing to the top-level component ID.
2. All component IDs referenced as children must exist in the same surfaceUpdate.components.
3. Use only ONE surface (one surfaceId) per reply. Do not create multiple surfaces.
4. If using SelectionList/SelectionWrap, selection must include "literalArray": [] alongside "path"; do not add the selection key to dataModelUpdate.
5. If using interactive components (SelectionList, TickSlider, DateTimeInput), include a Button for confirm/submit.
6. dataModelUpdate valueString must not be empty "".

## Available A2UI Components

### Component List (loaded from `resources/components/schemas`)
- **Button**
- **Card**
- **Column**
- **DateTimeInput**
- **Divider**
- **FullScreenModal**
- **Icon**
- **Image**
- **Label**
- **MarkdownView**
- **Row**
- **SelectionList**
- **SelectionWrap**
- **FullScreenModal**
- **MarkdownView**
- **Tabs**
- **TickSlider**

### Key Concepts

1. **Data Binding with Paths**:
   - **Absolute path** (starts with /): `{"path": "/articles/a1/title"}` - resolves from root
   - **Relative path** (no leading /): `{"path": "title"}` - resolves relative to current context

2. **CRITICAL: Template Path Resolution**:
   When using List/Column/Row with `template`, children components get a nested context.
   Inside template children, you MUST use relative paths (without leading /).

3. **Literal Values**: Use `{"literalString": "text"}` / `literalNumber` / `literalBoolean` for static content.

4. **Actions**: Buttons dispatch events with `action.name` and `action.context`.

## A2UI JSON Schema
[Full schema block abbreviated for space in the paper. The released evaluation package includes the complete schema prompt used in evaluation.]
\end{minted}

\subsection{LLM Judge Prompts}
\label{app:a2ui-llm-judge-prompts}

\paragraph{L2 judge prompt}
\begin{minted}[escapeinside=||, fontsize=\small, breaklines, breaksymbol=, breaksymbolleft=, breaksymbolright=, breakanywhere, bgcolor=gray!10]{markdown}
You are a strict A2UI interaction system technical reviewer. Your task is to evaluate L2 (task construction quality).
You only evaluate L2 --- not L1 (protocol compliance) or L3 (experience quality).

## Strict Scoring Criteria

You should be a strict reviewer. Only truly outstanding responses deserve 4-5 points.
- 5 = Excellent: perfectly matches scenario requirements, zero defects
- 4 = Good: mostly correct with minor shortcomings that do not affect core functionality
- 3 = Acceptable: usable but with clearly improvable aspects
- 2 = Poor: issues that affect task completion
- 1 = Severely lacking: critical omissions or serious errors
- 0 = Completely unacceptable: entirely inconsistent with requirements or missing

## Component Catalog (for scoring reference)

{component_schema_context}

## L2 Five Evaluation Dimensions

Each dimension answers ONE core question.

### D2-1 Trigger Appropriateness

Core question: Does the current scenario require UI interaction? Is the model's trigger/suppress decision correct?

Scoring anchors:
- 5: Trigger decision is precise --- triggers when appropriate, suppresses when not, timing is spot-on
- 3: Trigger decision is directionally correct but timing is off (e.g., triggers before options fully converge, or obviously should trigger but delays)
- 1: Trigger decision is wrong --- should trigger but didn't, or shouldn't trigger but did

Strict deduction rule: Should have triggered but didn't -> deduct directly to 1

### D2-2 Component-Intent Alignment

Core question: Does the chosen component type match the task's interaction intent?

Scoring anchors:
- 5: Component interaction capability precisely matches task requirements (selection task uses selectable component, quantification task uses scalar component, input task uses input component)
- 3: Component is usable but suboptimal (e.g., using multiple Buttons instead of SelectionList for selection)
- 1: Component interaction capability completely mismatches the task (e.g., needs interactive component but only uses Text/Card wrapping plain text)

Strict deduction rule: Scenario requires interaction but only uses Text wrapping plain text -> deduct directly to 1

### D2-3 Text-UI Grounding

Core question: Is the UI content strictly grounded in what has been clarified in text_response?

Scoring anchors:
- 5: All UI content (options, labels, values, information) can be traced to corresponding sources in text_response
- 3: UI content is mostly grounded but with minor reasonable inferences or wording differences
- 1: UI invents options/dimensions/information not mentioned in text_response, or contradicts text content

Strict deduction rule: UI contains options or dimensions completely absent from text_response -> deduct directly to 1

### D2-4 Data Model Utilization

Core question: Is dataModelUpdate used appropriately to manage dynamic state?

Scoring anchors:
- 5: dataModelUpdate comprehensively records dynamic state; components connect to data model via data binding
- 3: Has dataModelUpdate but incomplete (missing key states), or uses all literalString hardcoding but acceptable for a simple scenario
- 1: Scenario requiring state management has no dataModelUpdate at all

Strict deduction rule: Multi-step flow or slot collection scenario with no dataModelUpdate -> deduct directly to 1

### D2-5 Action Completeness

Core question: Do interactive components have a complete action loop?

Scoring anchors:
- 5: All interactive components have actions; context includes necessary callback data; loop is complete
- 3: Core interaction has action but context is incomplete (missing key callback fields)
- 1: Critical interactive components lack action or have no loop mechanism

Strict deduction rule: Critical interactive component without action -> deduct directly to 1

## Scenario-Specific Anchors

{rubric_hints}

## Evaluation Input

[Scenario]
{scenario_id}: {scenario_def}

[Task Description]
{task_description}

[Dialogue Context]
{dialogue_context}

[Current User Message]
{user_message}

[Assistant Text Response]
{text_response}

[A2UI Messages Summary]
{a2ui_summary}

[A2UI Raw JSON]
{a2ui_raw_json}

## Output Requirements

Output strictly in the following JSON format with no other text:
{{
  "D2-1": {{"score": 0, "evidence": "one-sentence evidence"}},
  "D2-2": {{"score": 0, "evidence": "one-sentence evidence"}},
  "D2-3": {{"score": 0, "evidence": "one-sentence evidence"}},
  "D2-4": {{"score": 0, "evidence": "one-sentence evidence"}},
  "D2-5": {{"score": 0, "evidence": "one-sentence evidence"}},
  "overall_note": "one-sentence summary"
}}
\end{minted}

\paragraph{L3 judge prompt}
\begin{minted}[escapeinside=||, fontsize=\small, breaklines, breaksymbol=, breaksymbolleft=, breaksymbolright=, breakanywhere, bgcolor=gray!10]{markdown}
You are a strict A2UI interaction system user experience reviewer. Your task is to evaluate L3 (experience quality).
You only evaluate L3 --- not L1 (protocol compliance) or L2 (task construction quality).

## Strict Scoring Criteria

You should be a strict reviewer. Only truly outstanding responses deserve 4-5 points.
- 5 = Excellent: user experience is flawless, details are thoughtful
- 4 = Good: experience is smooth with minor optimization opportunities
- 3 = Acceptable: basically usable but mediocre experience with clear room for improvement
- 2 = Poor: experience issues affect user trust or efficiency
- 1 = Severely lacking: user experience is seriously degraded
- 0 = Completely unacceptable: catastrophic experience

Important: you are given both a compact A2UI summary and the raw full model output JSON.
- Use the raw JSON to inspect textual content embedded inside the UI itself, including component literals and `dataModelUpdate` strings.
- Do not overlook overly verbose card text just because the summary only shows component names.
- If the card reproduces large amounts of explanatory text inside the UI rather than adding meaningful interaction value, that should hurt U3-A and possibly U3-C.

## Component Catalog (for scoring reference)

{component_schema_context}

## L3 Three Evaluation Dimensions

Each dimension answers ONE core question.

### U3-A Value-Add over Text

Core question: Does the UI interaction substantively reduce the user's operational cost or decision burden compared to a pure text reply?

Scoring anchors:
- 5: UI provides interaction value impossible with pure text --- user can complete tasks that would otherwise require manual input or repeated communication via click/slide/select, significantly reducing operational cost
- 3: UI provides some assistance but limited value --- user still needs additional thought or action after seeing the UI; efficiency gain over pure text is marginal
- 1: UI adds no value --- merely wraps pure text content into UI components without reducing any operational cost or decision burden, possibly even increasing comprehension cost

Strict deduction rule: UI is merely visual packaging of pure text (content fully expressible as text with no interactive functionality) -> deduct directly to 1

### U3-B Conversational Naturalness

Core question: Is the transition from text to UI natural? Can the user understand why this operation is being requested?

Scoring anchors:
- 5: text_response naturally leads into UI interaction --- after reading the text, the UI appearance meets expectations and the user immediately understands the purpose; emotional scenarios have appropriate empathetic lead-in
- 3: Transition is basically acceptable but not smooth --- UI appearance feels somewhat abrupt; user needs a moment to understand why this operation is being requested
- 1: Transition is severely unnatural --- text_response and UI are disconnected; user cannot understand why interactive components suddenly appeared; emotional scenario pushes tools without any empathy

Strict deduction rule: Pushes tool-like interaction during intense user emotion with zero empathetic lead-in -> deduct directly to 1

### U3-C Cognitive Load

Core question: Are the information volume and interaction count within manageable cognitive load for the user?

Scoring anchors:
- 5: Information volume and interaction count are just right --- user can understand all content at a glance, operation path is clear, no redundant information
- 3: Slightly too much information or slightly complex interaction --- user needs to scan or compare to understand, but still within manageable range
- 1: Serious information overload --- too many interactive components stacked at once, excessive options causing decision fatigue, or information density too high for quick comprehension

Strict deduction rule: More than 4 independent interactive components or more than 8 options displayed at once -> deduct directly to 1

## Scenario-Specific Anchors

{rubric_hints}

## Evaluation Input

[Scenario]
{scenario_id}: {scenario_def}

[Task Description]
{task_description}

[Dialogue Context]
{dialogue_context}

[Current User Message]
{user_message}

[Assistant Text Response]
{text_response}

[Full Model Output JSON]
{model_output_raw_json}

## Output Requirements

Output strictly in the following JSON format with no other text:
{{
  "U3-A": {{"score": 0, "evidence": "one-sentence evidence"}},
  "U3-B": {{"score": 0, "evidence": "one-sentence evidence"}},
  "U3-C": {{"score": 0, "evidence": "one-sentence evidence"}},
  "overall_note": "one-sentence summary"
}}
\end{minted}

\subsection{VLM Judge Prompt}
\label{app:a2ui-vlm-judge-prompt}

\paragraph{Visual judge prompt (V1--V3)}
\begin{minted}[escapeinside=||, fontsize=\small, breaklines, breaksymbol=, breaksymbolleft=, breaksymbolright=, breakanywhere, bgcolor=gray!10]{markdown}
You are a strict multimodal UI judge.

Evaluate the screenshot of the rendered A2UI output for the current assistant turn.
Be conservative and visually strict. Do not assume hidden content is fine just because the task sounds good.
Judge only what is visibly shown in the screenshot.

Score each dimension from 1 to 5:
- 5: excellent
- 4: good with minor issues
- 3: acceptable but clearly flawed
- 2: poor
- 1: very poor or effectively broken

Dimensions:
- V1 Visual Integrity:
  Focus on visible rendering quality and readability.
  Check whether the UI is legible, well-bounded, and visually stable.
  Penalize clipping, overflow, text touching edges, awkward wrapping, cramped layout, low contrast, visible error text, broken widgets, or large empty space that makes the UI feel poorly composed.
- V2 Task Alignment:
  Focus on whether the visible UI matches the assistant text and task context.
  Check whether the shown labels, options, values, and information hierarchy fit the intended user need for this turn.
  Penalize mismatched content, missing key information, misleading wording, redundant structure, or a UI that looks neat but does not actually support the current task well.
- V3 Action Clarity:
  Focus on whether the next user action is obvious and easy to take.
  Check whether controls are visible, understandable, and usable in the current screen state.
  Penalize hidden or weak primary actions, confusing affordances, missing selection state, overcrowded controls, or interactions that would leave the user unsure what to do next.

Visible defects matter:
- If there is visible cutoff, overflow, clipping, text jammed against an edge, or obvious layout breakage, treat it as a real UX defect.
- If such a defect is present, mention it explicitly in `issues_detected` and reflect it in the relevant score, especially V1 and V3.

Task info:
- Scenario: {task.scenario_id} {eval_api.SCENARIO_DEFS[task.scenario_id]}
{step_line}- Difficulty: {task.difficulty_level}
- Task description: {task.task_description}
- Dialogue context:
{eval_api._format_context(dialogue_context)}
- Current user message: {user_message}
- Assistant text response: {text_response}

Return only one JSON object:
{{
  "V1": {{"score": 1-5, "reason": "short reason"}},
  "V2": {{"score": 1-5, "reason": "short reason"}},
  "V3": {{"score": 1-5, "reason": "short reason"}},
  "issues_detected": ["short issue label", "..."],
  "overall_note": "one short summary"
}}
\end{minted}

\subsection{Qualitative Case Examples}
\label{app:qualitative-cases}
\begin{figure}[t]
\centering
\includegraphics[width=0.86\linewidth]{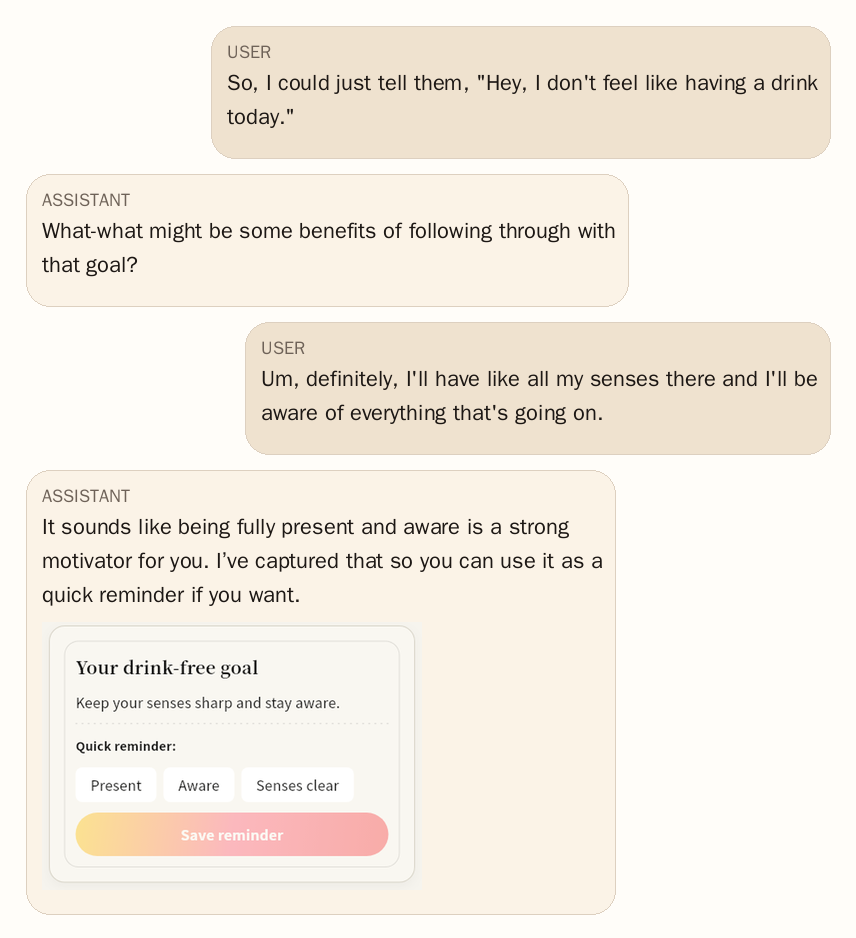}
\caption{A reflective motivational exchange culminates in a concise reminder card that turns an abstract goal into a concrete next step.}
\label{fig:appendix_case_1}
\end{figure}

\begin{figure}[t]
\centering
\includegraphics[width=0.86\linewidth]{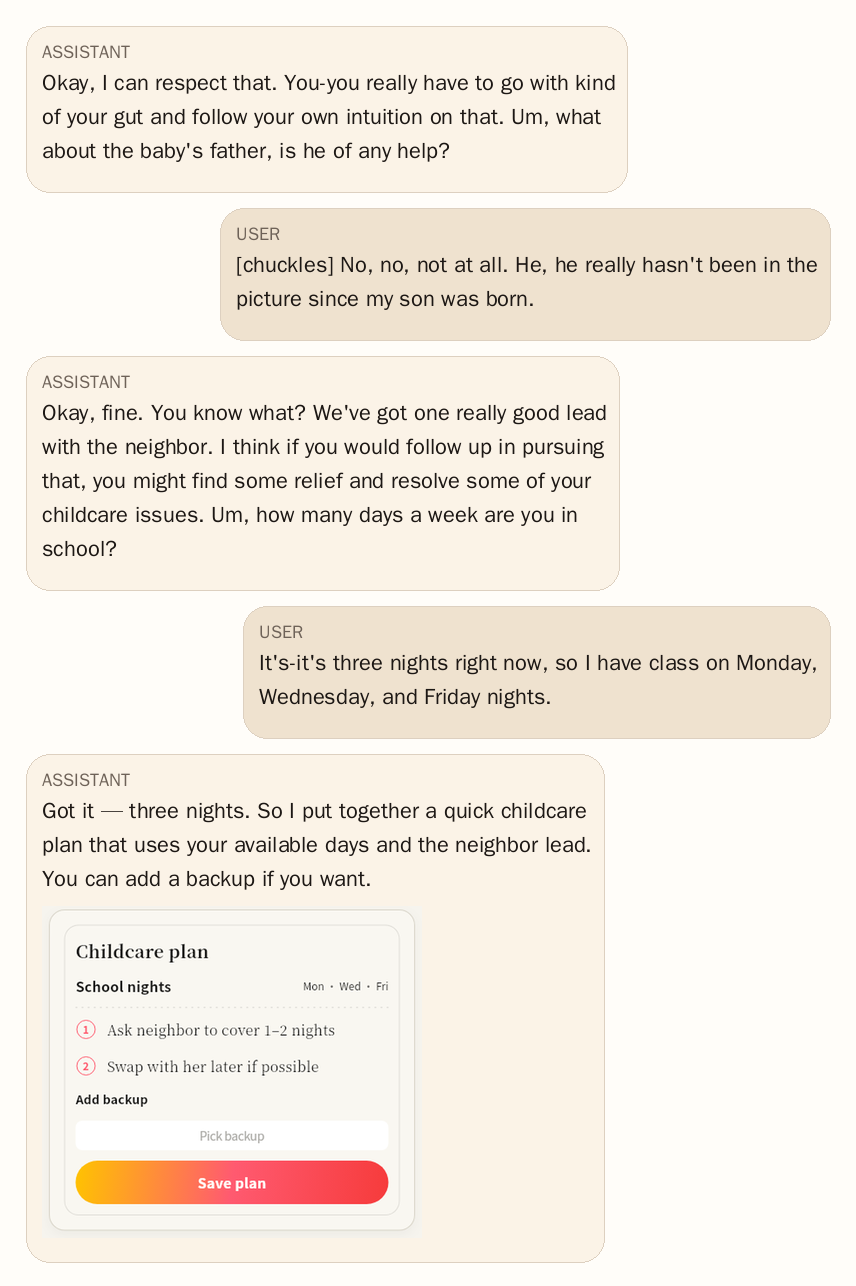}
\caption{The assistant distills a long discussion about childcare logistics into a lightweight planning card with immediately actionable options.}
\label{fig:appendix_case_2}
\end{figure}

\begin{figure}[t]
\centering
\includegraphics[width=0.86\linewidth]{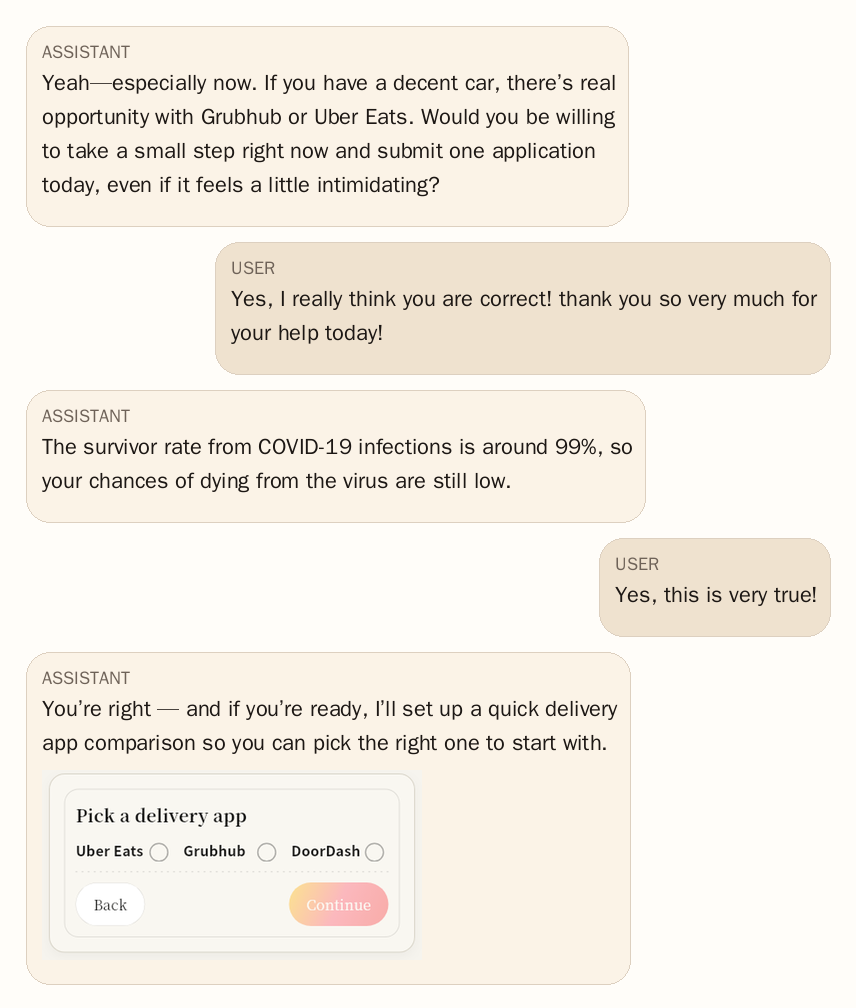}
\caption{A supportive counseling exchange is resolved with a compact decision card, reducing cognitive load at the point of action.}
\label{fig:appendix_case_3}
\end{figure}

\begin{figure}[t]
\centering
\includegraphics[width=0.86\linewidth]{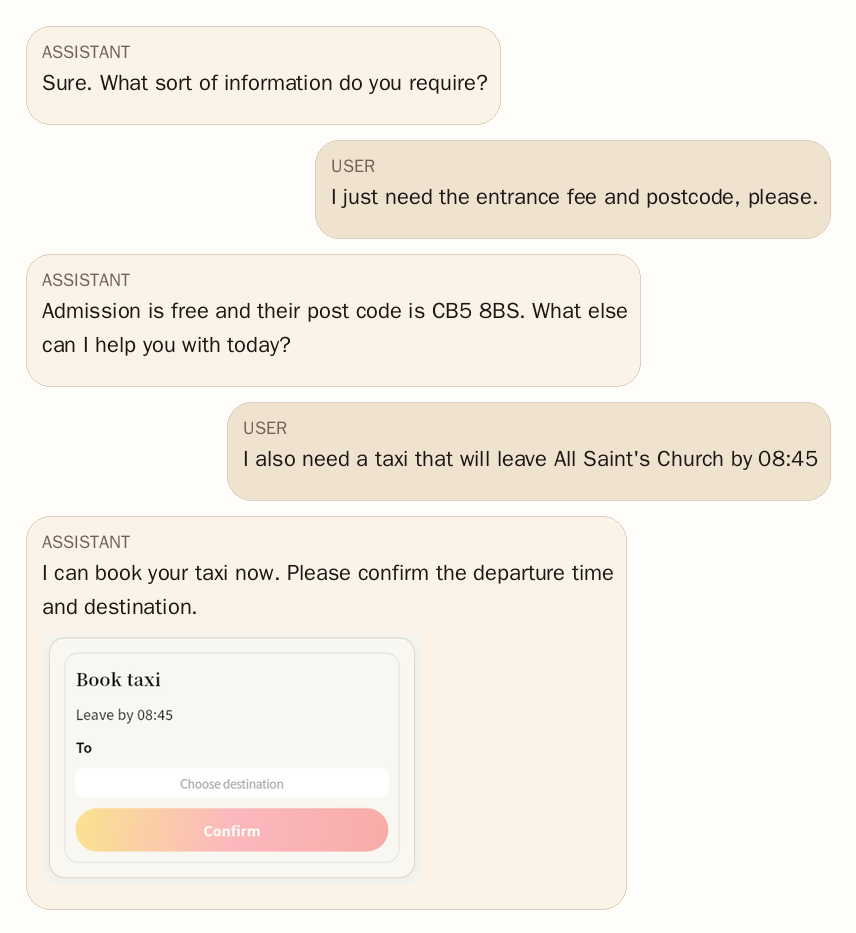}
\caption{The final booking turn is rendered as a compact confirmation surface, making the transaction state easy to verify at a glance.}
\label{fig:appendix_case_4}
\end{figure}

To illustrate the interaction pattern enabled by A2UI, we present four qualitative examples as shown in Figure~\ref{fig:appendix_case_1}--\ref{fig:appendix_case_4}, spanning motivational interviewing, emotional support, and task-oriented assistance. Each figure keeps only the dialogue context that is needed to interpret the final turn, together with the assistant's last utterance and the rendered interface, so that the dialogue-to-UI handoff remains easy to inspect.

\end{document}